\documentclass[10pt,journal]{IEEEtran}

\usepackage{amsmath,amssymb,amsfonts}
\usepackage{graphicx}
\usepackage{textcomp}
 \usepackage{algorithm, algorithmic}
 \usepackage{colortbl}
\usepackage{epsfig}
\usepackage{epstopdf}
\usepackage{enumerate}
\usepackage{diagbox}
\usepackage{subfigure}
\usepackage{cuted}
\usepackage{bm}
\usepackage{fancybox}
\usepackage{balance}

\usepackage{graphicx} 
\usepackage{graphics} 
\usepackage{amsmath} 
\usepackage{amssymb}  
\usepackage{xcolor}

\newcommand{\RR}{\mathbb{R}}

\newcommand{\WW}{\mathbb{W}}

\newcommand{\zero}{\mathbf{0}}
\newcommand{\G}{\mathcal{G}}

\newcommand{\E}{\mathcal{E}}
\newcommand{\LL}{\mathcal{L}}

\newenvironment{Proof}{\noindent{\em Proof:\/}}{\hfill $\Box$\par}
\newtheorem{Theorem}{Theorem}
\newtheorem{Lemma}{Lemma}

\newtheorem{Assumption}{Assumption}

\usepackage{enumerate}

\newtheorem{Remark}{Remark}

\newtheorem{Problem}{Problem}

\allowdisplaybreaks

\begin{document}
\title{\bf Cooperative Control of Parallel Actuators for Linear Robust Output Regulation of Uncertain Linear Minimum-phase Plants}
\author{Liang Xu, Tao Liu, and Zhiyun Lin,~\IEEEmembership{Fellow, IEEE}
  \thanks{L. Xu is with School of Mathematics and Statistics, Fuzhou University, Fuzhou 350116, China. E-mail: {\tt xul6@fzu.edu.cn}.}
  \thanks{T. Liu and Z. Lin are with Guangdong Provincial Key Laboratory of Fully Actuated System Control Theory and Technology, School of Automation and Intelligent Manufacturing, Southern University of Science and Technology, Shenzhen 518055, China. E-mail: {\tt liut6@sustech.edu.cn; linzy@sustech.edu.cn}. }}
\maketitle

\begin{abstract}
This paper investigates the robust output regulation problem for an uncertain linear minimum-phase plant with cooperative parallel operation of multiple actuators. 
Building on the internal model approach, we first propose a dynamic output feedback control law to solve the robust output regulation problem with a single actuator. 
Then, we construct a distributed dynamic output feedback control law 
that is nearly independent of the number of actuators and incorporates coupling terms 
to address the linear robust output regulation problem with cooperative parallel operation of multiple actuators over undirected communication networks.
We reveal the connection in the design of parameters between the dynamic output feedback control law under single actuator operation and the distributed dynamic output feedback control law under cooperative parallel operation with multiple actuators. Moreover, we remove the existing assumption that the actuator dynamics must be Hurwitz stable, thereby enabling the incorporation of unstable actuators in our framework.
Finally, two numerical examples are provided to validate the effectiveness of the proposed control laws.
\end{abstract}

\begin{IEEEkeywords} 
Robust output regulation, cooperative parallel operation, 
multiple actuators, linear minimum-phase systems.
\end{IEEEkeywords}

\section{Introduction}
Tracking and disturbance rejection are fundamental challenges in systems and control theory. 
When reference trajectories and disturbances are generated by an autonomous system 
called an exosystem, the output tracking problem and the disturbance rejection problem
can be tackled simultaneously within the classical output regulation framework. 
This framework formulates tracking and disturbance rejection 
as a unified output regulation problem, 
providing a systematic solution applicable to both certain and uncertain plants, see, e.g., \cite{Davison-1976TAC}, \cite{Francis-Wonham-1976Auto}, and \cite{Francis-1977SIAM}. 
Within this framework, two classical approaches have been developed 
to solve the output regulation problem for linear systems: 
the feedforward control approach, which relies on accurate plant models, 
and the internal model approach, which handles uncertain plant models. 
Over the years, output regulation has attracted substantial research interest, 
leading to various extensions such as 
nonlinear output regulation~\cite{Chen-Huang-2015book}, \cite{Huang2004}, 
event-triggered output regulation \cite{Su_Xu_Wang_Xu-2019Auto}, \cite{Wang-Shu-Chen-2023Auto}, 
and cooperative output regulation \cite{Cai-Su-Huang-2022book}, \cite{DengZhangFeng2022}.

However, for classical output regulation problems, most existing control formulations implicitly assume instantaneous and ideal actuation, neglecting the dynamic characteristics between the controller output and the plant input.
This conventional configuration fails to address two important considerations in practical control systems.
First, the dynamics of essential actuators, such as inverters, DC-DC/AC-DC converters, and electric motors,
can have significant impact on system performance and thus should be explicitly accounted for in control design.
A recent work by~\cite{Verrelli_Tomei-2023TAC}
has begun addressing this gap by solving the output regulation problem for an unknown stable linear plant with an uncertain actuator, 
where the input to the plant is generated by a minimum-phase linear actuator. In addition, reference \cite{He-Huang-2025IJRNC} studied the exponential tracking and disturbance rejection problem of a class of Euler–Lagrange systems with high-order actuator dynamics.
Second, traditional feedback control methods are considered passive, 
which can lead to limitations on both performance and implementation \cite{Wei_Li_Wang-2020CAA}. 
To overcome these limitations, 
the innovative parallel control methodology 
was developed by \cite{Wang-2006} and \cite{Wei_Li_Wang-2020CAA}.
The core idea of this approach is to extend practical control problems into a virtual space, 
enabling enhanced control through virtual reality interactions. 
Subsequent research has greatly advanced parallel control 
for tracking and output regulation problems.
Reference~\cite{Lu_Wei_Wang-2020CAA} introduced an adaptive dynamic programming approach 
to optimal tracking for nonlinear systems by parallel control, 
which maintains input continuity even when the reference signals contain finite jump discontinuities. 
This work was later generalized in \cite{Wei_Jiao_Dong-2024TCYB}
to handle systems with parametric uncertainties. 
Further developments by \cite{Wei_Li_Wang-2023TCYB} 
established necessary and sufficient conditions for the existence of parallel regulators 
in linear output regulation with external disturbances. 

Though some recent studies have explicitly considered actuator dynamics in output regulation, whether involving a real   or ``virtual'' actuator\footnote{In this paper, we refer to a system that can reconstruct the state of the parallel controller as a ``virtual'' actuator. For instance, in the framework of parallel control proposed in \cite{Wei_Li_Wang-2020CAA}, a parallel controller can be described by $\dot{u} = g(x, u)$. The state $u$ of this parallel controller can be viewed as the output of the following virtual actuator:
	\begin{align*} 
		\dot{z} = g_1(z, v), \qquad y = z 
	\end{align*} 
which is controlled by $v = g_2(x, z)$, with $g_1(z, g_2(x, z)) = g(x, z)$ and $u(0) = z(0)$.
}, they assumed that the plant is driven by a single actuator. 
Nevertheless, the configuration with a single actuator  
often suffers from inadequate actuation capacity, 
particularly when dealing with complex, large-scale, and heavy control systems. 
In practical engineering applications, 
this inadequacy is commonly addressed through a parallel configuration of multiple actuators,
which not only provides greater actuation capacity but also enhances system redundancy and reliability through cooperative parallel operation. 
Therefore, it is interesting to study 
the output regulation problem where the plant is cooperatively driven by parallel actuators.

There has been some research on parallel operation of specific actuators
in the literature, such as inverters, DC-DC converters, and electric motors, 
with applications to DC microgrids, power systems, and motor-driven systems, respectively. 
The two primary objectives of parallel operation of multiple actuators are output regulation and input sharing, and the specific regulated output and shared input vary across different types of actuators. For instance, parallel DC-DC converters require voltage regulation and current sharing \cite{Sadabadi-2021CSL}, while parallel electric motors need speed regulation and load torque sharing~\cite{Galassini-Menis-2018TIE}. 
To achieve these objectives, various control strategies have been developed, 
which can be categorized as centralized, decentralized, or distributed.
Primary-secondary control is a typical centralized control strategy 
 and has been used in \cite{Chen_Chu-1995TPE} and \cite{Siri_Lee_Wu-1992TAES} 
 for parallel inverter and converter systems, respectively. 
Although effective for input sharing, 
 this approach exhibits limited redundancy and 
 remains vulnerable to single-point failures.
 Droop control is a decentralized control strategy that achieves input sharing without communication links, making it popular for primary control of DC microgrids \cite{Zhang_Schiffer-2022Auto}, \cite{Li_Chan_Hu_Guerrero-2021TIE}. 
However, its simplicity comes at the cost of slow response and reduced regulation accuracy \cite{Hu_Shan_Cheng_Islam-2022TPE}. 
Further, distributed control offers a superior alternative by incorporating communication networks that enable information exchange among neighboring subsystems. 
This approach improves both system reliability and control performance.  
For example, a consensus-based distributed control scheme was developed in \cite{Lee_Oh-2021IJCAS} for motor-driven systems to ensure precise speed regulation 
and equal load torque sharing.
Similarly, distributed control has been applied to achieve voltage regulation 
and current sharing in parallel DC-DC converters \cite{Sadabadi-2021CSL}.
In addition, distributed control strategies have been widely adopted 
in secondary control of DC microgrids to compensate for primary droop control errors 
while avoiding the single-point failure risk of centralized control strategies; see, e.g., \cite{Lou_Gu_Hong-2020TSG}, \cite{Qu_Liu_Tian-2024TIE}, and \cite{Zhao_Yang_Lai-2025TSG}.

In contrast to the application-specific results discussed above, 
Lim and Oh \cite{Lim_Oh-2024Auto} recently established 
a systematic framework for cooperative parallel operation of multiple actuators, 
where both the plant and the actuators are described by precise linear system models. 
We subsequently extended their framework 
to develop a robust version that is capable of handling uncertain linear plants \cite{Xu_Su_Liu-2025Auto}. However, these two frameworks present several notable limitations.
First, both approaches in \cite{Lim_Oh-2024Auto} and \cite{Xu_Su_Liu-2025Auto}  rely on the Hurwitz stability assumption for the actuators, which excludes common integrator-type actuators, such as reaction/momentum wheels \cite{Markley-Crassidis-2014book} and elastic-joint robot actuators \cite{He-Huang-2025IJRNC}.
Second, the control law design procedures in both \cite{Lim_Oh-2024Auto} and \cite{Xu_Su_Liu-2025Auto} are inherently dependent on the number of actuators. 
As a result, any change in the number of actuators requires a complete redesign of the control law, thereby impacting scalability and implementation. 
Third, while the robust version in \cite{Xu_Su_Liu-2025Auto} demonstrates some tolerance to uncertainties, 
the admissible uncertainty range cannot be determined a priori, 
hence limiting practical applicability.

In this paper, we aim to develop a simple, scalable, and robust framework for solving the linear robust output regulation problem with cooperative parallel operation of multiple actuators. 
Specifically, to address the need of prescribing the range of uncertainties, 
we focus on the robust output regulation problem for a class of uncertain linear  minimum-phase plants. The main contributions of this paper are threefold:
\begin{itemize}

	\item First, we develop a novel distributed dynamic output feedback control law that achieves almost complete independence from the number of the actuators over undirected communication networks.

	\item Second, we uncover a fundamental linkage connecting robust output regulation in the context of single actuator operation and cooperative parallel operation with multiple actuators.

   \item Third, we remove the restrictive Hurwitz stability requirement, 
   extending our approach to handle not only stable actuators but also those with marginally stable 
   or potentially unstable dynamics.

\end{itemize}
More explicitly, building on the internal model principle, 
we first propose a dynamic output feedback control law to solve 
the linear robust output regulation problem for the special case 
where the plant is driven by a single actuator. 
Next, by analyzing the sum dynamics of the control inputs and internal models, we derive a fundamental connection between robust output regulation under single actuator operation and that under cooperative parallel operation with multiple actuators.
We then develop a distributed dynamic output feedback control law
where only a single gain parameter depends on the number of actuators. 
This distributed control law ensures not only output regulation of the closed-loop system,
but also equal sharing of the plant input among the actuators.
Unlike previous approaches in \cite{Lim_Oh-2024Auto} and \cite{Xu_Su_Liu-2025Auto} that rely on the Hurwitz stability assumption on the actuators to guarantee the input sharing property, our approach attains this property through integrated consideration of actuator and internal model dynamics. This key insight allows us to eliminate the restrictive Hurwitz stability requirement via a proper parameter design.

The rest of this paper is organized as follows. 
In Section~\ref{Section-Problem formulation}, 
we formulate the linear robust output regulation problem 
with cooperative control of parallel actuators. 
A special case, namely, the linear robust output regulation problem 
with a single actuator is investigated in Section~\ref{section-single-results}. In Section~\ref{Section-MR}, we demonstrate the solvability of 
the linear robust output regulation problem 
with cooperative control of parallel actuators over undirected communication networks. 
Section~\ref{Section-Example} presents two numerical examples to illustrate our designs and Section~\ref{Section-Conclusions} concludes the paper.

\medskip
\noindent
\textit{Notation}: 
$\RR$ denotes the set of real numbers.
For $x_{i}\in \RR^{n_{i}}, i=1, \ldots,N$, $\mathrm{col}(x_{1},\ldots,x_{N})=\begin{bmatrix} x_{1}^{T} & \cdots & x_{N}^{T} \end{bmatrix}^{T}$.
$\zero $ denotes a matrix of zeros with appropriate dimensions.
$\mathbf{1}_{N}$ denotes an $N$-dimensional column vector whose coordinates are all $1$.  
The symbol $\otimes$ represents the Kronecker product of matrices.

\section{Problem Formulation}\label{Section-Problem formulation}
Consider an uncertain linear plant as follows:
\begin{align}\label{eq-plant}
	\dot z & =A_1(w) z+A_2(w) \xi_1+E_0(w) v \notag\\
	\dot{\xi}_s & =\xi_{s+1}, \quad s=1, \ldots, r-1 \notag\\
	\dot{\xi}_r & =A_3(w) z+\sum_{s=1}^r c_s(w) \xi_s+E_r(w) v+b(w) u_p \notag\\
	y & =\xi_1
\end{align}
where $z\in \RR^{n-r}$ and $\xi=\mathrm{col}(\xi_1, \dots, \xi_r)\in\RR^{r}$ are the states, $u_p\in\RR$ and $y\in\RR$ are the input and output, respectively; 
$w\in\WW$ is an uncertain parameter vector with $\WW\subseteq\RR^{n_w}$ being a compact set; 
$v\in\RR^q$ is the exogenous signal representing the reference input and/or the external disturbance 
and is assumed to be generated by a linear exosystem as follows:
\begin{align}\label{eq-exosystem}
	\dot v & =Sv \notag \\
   y_{0}&=F(w)v
\end{align}
where $S\in \RR^{q\times q}$ is the system matrix and $y_{0} \in \RR$ is the output. 
The regulated error is defined as $e=y-y_{0}$. In the plant \eqref{eq-plant} and the exosystem \eqref{eq-exosystem}, $A_1(w)\in\RR^{(n-r)\times (n-r)}$, $A_2(w)\in\RR^{(n-r)\times 1}$, $A_3(w)\in\RR^{1\times(n-r)}$, $E_0(w)\in\RR^{(n-r) \times q}$, $E_r(w)\in\RR^{1 \times q}$, and $F(w)\in\RR^{1\times q}$ are uncertain matrices, 
$c_{s}(w), s=1,\dots, r$, and $b(w)$ are uncertain constants, 
and they all depend continuously on the uncertain parameter vector $w$.

In this paper, we assume the plant \eqref{eq-plant} is minimum-phase, which implies that $A_{1}(w)$ is Hurwitz for all $w\in\WW$. Moreover, we assume  the control direction is known, i.e., $b(w)>0$ for all $w\in\WW$.

Suppose that the plant \eqref{eq-plant} is driven by $N$ linear actuators  as follows:
	\begin{align}\label{eq-multiple-actuators}
		\dot x_i &=a x_i+b_au_i \notag\\
		y_i &=x_i, \quad i=1,\dots, N \notag\\
		u_p&=\sum_{i=1}^{N}y_i 
	\end{align}
where $x_i\in \RR$, $u_i\in \RR$, and $y_i\in \RR$ are the state, control input, and output of the $i$th actuator, respectively; $u_{p}\in \RR$ is the sum of the outputs of all actuators and also the input to the plant, $a\in \RR$ and $b_a>0$ are two known constants.

For each actuator in \eqref{eq-multiple-actuators},
a controller will be designed.
Moreover, it is assumed that these controllers can exchange information over a communication network. 
The communication network among the $N$ controllers is modeled as an undirected graph $\mathcal{G} = (\mathcal{V}, \mathcal{E})$, where $\mathcal{V} = \{1, \ldots, N\}$ represents the controllers 
and $\mathcal{E} \subseteq \mathcal{V} \times \mathcal{V}$ represents the communication links. 
In particular, an edge $(j, i) \in \mathcal{E}$ indicates that the $i$th and $j$th controllers can exchange  information with each other.

We will consider designing a distributed dynamic output feedback control law 
comprising the $N$ controllers in the following form:
\begin{align}\label{eq-control-law-general-form}
	u_i&=K\varrho_i, \quad i=1,\dots,N \notag\\
	\dot\varrho_i&=H\left(e,y_i,\varrho_i, \{\varrho_j-\varrho_i:j\in\mathcal{N}_i\}\right)
\end{align}
where, for $i=1,\dots, N$, $\mathcal{N}_i=\{j \in \mathcal{V} : (j,i)\in\E\}$ is the neighbor set of the $i$th controller,
 $\varrho_i\in\RR^{n_\varrho}$ is the state of the $i$th dynamic compensator with 
 the dimension $n_\varrho$ to be specified, $K\in\RR^{1\times n_\varrho}$ is a constant row vector, 
 and $H(\cdot)$ is a linear function of its arguments.

We now present the problem of cooperative control of parallel actuators for robust output regulation
of an uncertain linear minimum-phase plant.

\begin{Problem}\label{Problem-Robust_output_regulation_multiple_actuators}
	Given the plant \eqref{eq-plant}, the exosystem \eqref{eq-exosystem}, the $N$ actuators in \eqref{eq-multiple-actuators}, a communication graph $\G$, 
and any compact set $\WW$ containing the origin, 
find a distributed dynamic output feedback control law of the form \eqref{eq-control-law-general-form}, such that for any $w\in\WW$, 
the following three properties are satisfied:
	\begin{enumerate}[(I)]
		\item The origin of the closed-loop system is asymptotically stable when $v=0$. \label{Property-one-multiple}
		\item For any initial conditions $z(0)$, $\xi(0)$, $v(0)$, and $x_{i}(0), \varrho_i(0), i=1,\dots,N$, the regulated error tends to zero, i.e., $\lim\limits_{t \to \infty} e(t)=0$.\label{Property-two-multiple}
		\item The input to the plant is shared by the $N$ actuators, i.e., $\lim_{t \to \infty}\left(y_{i}(t)-y_{j}(t)\right)=0, \forall\, i,j=1,\dots, N$.   \label{Property-three-multiple}
	\end{enumerate}
\end{Problem}

\begin{Remark}\label{Remark-comparsion-Problem-1}
Both classical robust output regulation and cooperative robust output regulation for uncertain linear minimum-phase systems have been studied in \cite{Liu-Huang-2017IJRNC}, \cite{Marino_Tomei-2021Auto}, \cite{Su-Huang-2014IJRNC}, \cite{Wu_Lu_Chen-2024IJRNC}, and \cite{Wu_Lu_Chen-2024TCNS}. 
Additionally, linear output regulation with a single (``virtual'') actuator has been explored in \cite{Verrelli_Tomei-2023TAC} and \cite{Wei_Li_Wang-2023TCYB}. 
The key distinction between Problem~\ref{Problem-Robust_output_regulation_multiple_actuators} 
and these priori works is the  explicit incorporation of cooperative control of multiple actuators, 
which was not previously addressed. 
Consequently, we must further ensure the additional Property~\eqref{Property-three-multiple} in Problem~\ref{Problem-Robust_output_regulation_multiple_actuators}
to account for parallel operation of multiple actuators.
\end{Remark}

\begin{Remark}
It is noted that cooperative parallel operation of multiple actuators for general linear systems 
was first studied in \cite{Lim_Oh-2024Auto} where precise knowledge of the plant is required. 
More recently, we extended the work in \cite{Lim_Oh-2024Auto} to a robust setting in \cite{Xu_Su_Liu-2025Auto},
allowing the plant to undergo certain uncertainties. 
However, the open set $\WW$ in \cite{Xu_Su_Liu-2025Auto}, 
which represents the set to which the uncertainties belong, 
cannot be explicitly quantified and thus is typically required to be sufficiently small.
In contrast, this paper considers a prescribed compact set  $\WW$ that can be arbitrarily large.
\end{Remark}

To address Problem \ref{Problem-Robust_output_regulation_multiple_actuators}, 
we make two assumptions.
One is standard in the output regulation literature \cite{Huang2004}
and the other is commonly used in the cooperative control literature \cite{Wu_Lu_Chen-2024IJRNC}.

\begin{Assumption}\label{Ass-nonnegative}
	The eigenvalues of $S$ have nonnegative real parts.
\end{Assumption}

\begin{Assumption}\label{Ass-connected-graph}
	The graph $\G$ is connected.
\end{Assumption}

\section{Robust Output Regulation with A Single Actuator }\label{section-single-results}
In this section, we first study a special case of Problem \ref{Problem-Robust_output_regulation_multiple_actuators},
where the plant \eqref{eq-plant} is driven by a single actuator as follows:
	\begin{align}\label{eq-actuator-single}
		\dot x_1 &=a x_1+b_au_1 \notag\\
		y_1&=x_1   \notag\\
		u_p&=y_1.
	\end{align}
It can be seen that \eqref{eq-actuator-single} is 
in the form of \eqref{eq-multiple-actuators} with $N=1$. 
Thus, in this special case,
Problem \ref{Problem-Robust_output_regulation_multiple_actuators} reduces 
to a linear robust output regulation problem with the control of a single actuator, 
as formulated below.

\begin{Problem}\label{Problem-Robust_output_regulation}
	Given the plant \eqref{eq-plant}, the exosystem \eqref{eq-exosystem}, the actuator \eqref{eq-actuator-single}, and any compact set $\WW$ containing the origin, 
find a dynamic output feedback control law such that for any $w\in\WW$,
the following two properties are satisfied: 
	\begin{enumerate}[(i)]
		\item The origin of the closed-loop system is asymptotically stable when $v=0$. \label{Property-one-single}
		\item For any initial conditions $z(0)$, $\xi(0)$, $v(0)$, and $x_1(0)$, the regulated error tends to zero, i.e., $\lim\limits_{t \to \infty} e(t)=0$. \label{Property-two-single}
	\end{enumerate}
\end{Problem}

\begin{Remark}
It is interesting to note that both references \cite{Verrelli_Tomei-2023TAC} and \cite{Wei_Li_Wang-2023TCYB} 
have studied linear output regulation involving a single (``virtual'') actuator.
In particular, reference \cite{Verrelli_Tomei-2023TAC} additionally requires the plant to be open-loop stable. In comparison, we consider the robust version of this problem for an uncertain linear minimum-phase plant in Problem~\ref{Problem-Robust_output_regulation} without requiring the internal stability of the plant. Such plants encompass many practical systems, including motor-driven systems, active suspension systems, and mass-spring-damper systems. Additionally, the actuator model \eqref{eq-actuator-single} can be naturally extended to generic non-Hurwitz stable actuators represented in controllable canonical form.
\end{Remark}
	
As established in \cite[Chapter 7]{Chen-Huang-2015book}, 
the internal model approach provides an effective solution to the robust output regulation problem.
This approach consists of two steps: 
\begin{itemize}
	\item Step 1: Design a dynamic compensator, called an internal model, that converts the robust output regulation problem for an uncertain system into a robust stabilization problem for an augmented system composed of the original system and the internal model.
	\item Step 2: Synthesize a stabilizing control law for the resulting augmented system.
\end{itemize}
Following this framework, in the rest of this section, 
we will construct an internal model and derive the augmented system. 
Then, we will demonstrate the solvability of Problem~\ref{Problem-Robust_output_regulation} 
through the development of a specific stabilizing control law for  the augmented system.

\subsection{Internal Model}\label{Section-Internal_Model}
Recall that $A_1(w)$ is Hurwitz for all $w \in \WW$. 
Under Assumption \ref{Ass-nonnegative}, the spectra of the matrices $S$ and $A_1(w)$ are disjoint. 
Thus, there exists a unique solution $Z(w) \in \RR^{(n-r) \times q}$ to the following Sylvester equation:
\begin{align*}
	Z(w)S=A_1(w) Z(w)+A_2(w) F(w)+E_0(w).
\end{align*}
Denote 
\begin{align*}
	\Pi(w)&=\mathrm{col}\left(F(w), F(w)S, \dots, F(w)S^{r-1}\right) \in \RR^{r \times q} \\
	\Xi(w)&=\frac{1}{b(w)}\Big(F(w) S^r -A_{3}(w) Z(w) \notag\\
	&\quad -\sum_{s=1}^r c_{s}(w) F(w) S^{s-1}-E_{r}(w)\Big) \in \RR^{1 \times q} \\
	U(w)&=\frac{1}{b_a}\left(\Xi(w) S-a \Xi(w)\right) \in \RR^{1 \times q}.
\end{align*} 
Then, $\mathrm{col}\left(Z(w)v, \Pi(w)v, \Xi(w)v\right)$ and $U(w)v$ represent the steady-state state of $\mathrm{col}(z, \xi, x_1)$ and the steady-state input of $u_1$, respectively, at which, $e = 0$.

Let $p(\lambda) = \lambda^l + \alpha_{l-1} \lambda^{l-1} + \cdots + \alpha_1 \lambda + \alpha_0$ be the minimal polynomial of $S$ and define
\begin{align}\label{eq-Phi-Psi}
	\Phi&=\left[\begin{array}{cccc}
		0 & 1 & \cdots & 0 \\
		\vdots & \vdots & \ddots & \vdots \\
		0 & 0 & \cdots & 1 \\
		-\alpha_0 & -\alpha_1 & \cdots & -\alpha_{l-1}
	\end{array}\right] \in \RR^{l \times l}\notag\\
 \Psi&=\begin{bmatrix}
         1 & 0 & \cdots & 0 
       \end{bmatrix}\in \RR^{1\times l}.
\end{align}
For $i=1,2$, select a controllable pair $(M_i,N_i)$ with $M_i\in\RR^{l\times l}$ being Hurwitz and $N_i\in\RR^{l\times 1}$. 
Since $M_i$ and $\Phi$ have no eigenvalues in common,
$(M_i,N_i)$ is controllable, and $(\Phi,\Psi)$ is observable, 
by \cite[Theorem 8.4]{Chen99}, 
each of the following Sylvester equations:
\begin{align}\label{eq-Sylvester_equation}
	T_i\Phi - M_iT_i = N_i\Psi, \quad i=1,2
\end{align}
 admits a unique solution $T_i \in \RR^{l\times l}$, which is nonsingular.

Define 
\begin{align*}
	\Theta_1(w,v)&=T_1\, \mathrm{col}\left(\Xi(w), \Xi(w)S, \dots, \Xi(w)S^{l-1}\right)v  \in \RR^{l}\\
	\Theta_2(w,v)&=T_2\, \mathrm{col}\left(U(w), U(w)S, \dots, U(w)S^{l-1}\right)v \in \RR^{l}.
\end{align*}
Then, it can be verified that
\begin{align}\label{eq-steady_state_generator_xa}
	\dot\Theta_1(w,v)&=T_1\Phi T^{-1}_1\Theta_1(w,v) \notag\\
	\Xi(w)v&=\Psi T^{-1}_1\Theta_1(w,v) 
\end{align}
and
\begin{align}\label{eq-steady_state_generator_u}
	\dot\Theta_2(w,v)&=T_2\Phi T^{-1}_2\Theta_2(w,v) \notag\\
	U(w)v&=\Psi T^{-1}_2\Theta_2(w,v).
\end{align}
In words, systems \eqref{eq-steady_state_generator_xa} and \eqref{eq-steady_state_generator_u}
are the steady-state generators for the steady-state state of $x_1$ and the steady-state input of $u_1$, respectively.
However, since the steady-state generators \eqref{eq-steady_state_generator_xa} and \eqref{eq-steady_state_generator_u} depend on the uncertain parameter vector $w$, they are not implementable. To circumvent this issue, we construct an internal model consisting of two asymptotic observers that estimate these steady-state generators as follows:
\begin{subequations}\label{Internal_Model}
  \begin{align}
	\dot\eta_1&=M_1\eta_1+N_1x_1   \\
	\dot\eta_2&=M_2\eta_2+N_2u_1.
\end{align}
\end{subequations}

\subsection{Augmented System}\label{Section-Augmented system}
Putting the plant \eqref{eq-plant}, the actuator~\eqref{eq-actuator-single}, and the internal model \eqref{Internal_Model} together gives the following augmented system:
\begin{align}\label{eq-augmented-system-single}
	\dot z & =A_1(w) z+A_2(w) \xi_1+E_0(w) v \notag\\
	\dot{\xi}_s & =\xi_{s+1}, \quad s=1, \ldots, r-1 \notag\\
	\dot{\xi}_r & =A_3(w) z+\sum_{s=1}^r c_s(w) \xi_s+E_r(w) v+b(w) x_1 \notag\\
	\dot{x}_{1} &=ax_{1}+b_a u_{1} \notag \\
	\dot\eta_1&=M_1\eta_1+N_1x_1  \notag\\
	\dot\eta_2&=M_2\eta_2+N_2u_1 \notag \\
	e & =\xi_1-F(w)v.
\end{align}
Perform the coordinate and input transformations below:
\begin{align*}
	\bar z & =z-Z(w)v  \notag\\
	\bar\xi_s&=\xi_s-F(w)S^{s-1}v, \quad s=1,\ldots,r \notag\\
	\bar x_1&=x_1-\Psi T^{-1}_1\eta_1 \notag\\
	\bar u_1&=u_1-\Psi T^{-1}_2\eta_2   \notag\\
	\bar\eta_1&=\eta_1-\Theta_1(w,v) \notag\\
	\bar\eta_2&=\eta_2-\Theta_2(w,v). 
\end{align*}
Then, the augmented system \eqref{eq-augmented-system-single} is transformed into
\begin{align}\label{eq-transformed-augmented-system}
	\dot{\bar z} & = A_1(w) \bar z+A_2(w) \bar{\xi}_1 \notag\\
	\dot{\bar\xi}_s & = \bar{\xi}_{s+1}, \quad s=1, \ldots, r-1  \notag\\
	\dot{\bar\xi}_r & = A_3(w) \bar{z}+\sum_{s=1}^r c_s(w) \bar{\xi}_s+b(w)\bar x_1+b(w)\Psi T^{-1}_1\bar\eta_1 \notag\\
	\dot{\bar x}_1 & = (a-\Psi T^{-1}_1N_1) \bar x_1+\Psi T^{-1}_1(aI_{l}-M_1-N_1\Psi T^{-1}_1)\bar \eta_1\notag\\
	&\quad+b_a\Psi T^{-1}_2\bar\eta_2+b_a \bar u_1 \notag\\
	\dot{\bar\eta}_1 & = \left(M_1+N_1\Psi T^{-1}_1\right) \bar\eta_1+N_1 \bar x_1  \notag\\
	\dot{\bar\eta}_2 & = \left(M_2+N_2\Psi T^{-1}_2\right) \bar\eta_2+N_2 \bar u_1  \notag\\
	e & =\bar\xi_1.
\end{align}
Further, inspired by \cite[equation (7.72)]{Chen-Huang-2015book}, we perform an additional coordinate transformation on system \eqref{eq-transformed-augmented-system} as follows:
\begin{align}\label{eq-transformation-2}
	\zeta_1 &=\bar\xi_r+\gamma_{r-2} \bar\xi_{ r-1}+\cdots+\gamma_1 \bar\xi_2+\gamma_0 \bar\xi_1 \notag\\
	\zeta_2 &=\bar x_1+k_1\zeta_1    \notag \\
	\tilde\eta_1 &=\bar\eta_1-\frac{1}{b(w)} N_1 \zeta_1 \notag\\
	\tilde\eta_2 &=\bar\eta_2-\frac{1}{b_a} N_2 \zeta_2
\end{align}
where the constants $\gamma_{s}, s=0,\dots, r-2$, are selected such that the polynomial 
\begin{align}\label{eq-stable_polynomial}
	f(\lambda)=\lambda^{r-1}+\gamma_{r-2}\lambda^{r-2}+\dots+\gamma_{1}\lambda+\gamma_{0}
\end{align}
is stable and $k_1$ is a positive gain to be determined.
Then, the augmented system \eqref{eq-transformed-augmented-system} is transformed into 
\begin{align}\label{eq-augmented_system_transformed}
	\dot{\bar z} & = A_1(w) \bar z+A_2(w)D \bar\xi   \notag\\
	\dot{\bar\xi}& =\Lambda\bar\xi+G \zeta_1   \notag\\
	\dot{\zeta}_1 & =A_3(w) \bar{z}+C(w)\bar\xi+\tilde c(w,k_1)\zeta_1+b(w)\Psi T^{-1}_1\tilde\eta_1 \notag\\
	&\quad\ +b(w)\zeta_2    \notag\\
	\dot{\tilde \eta}_1 & =M_1\tilde\eta_1+\check{A}_3(w)\bar z+\check {C}_1(w)\bar\xi+\check{C}_2(w)\zeta_1 \notag\\
	\dot\zeta_2 & =\bar A_3(w,k_1) \bar{z}+\bar C_1(w,k_1)\bar\xi+\bar c_1(w,k_1)\zeta_1+\bar C_2(w,k_1)\tilde\eta_1 \notag\\
	&\quad\ +\bar c_2(w,k_1)\zeta_2+b_a\Psi T^{-1}_2\tilde\eta_2+b_a\bar u_1 \notag\\
	\dot{\tilde \eta}_2 & =M_2 \tilde \eta_2+\hat A_3(w,k_1)\bar z+\hat C_1(w,k_1)\bar\xi+\hat C_2(w,k_1)\zeta_1 \notag\\
	&\quad\ +\hat C_3(w,k_1)\tilde\eta_1+\hat C_4(w,k_1)\zeta_2 
\end{align}
where $\bar\xi=\mathrm{col}\left(\bar\xi_1, \dots, \bar\xi_{r-1}\right) \in \RR^{r-1}$,
\begin{align*}
D&=\begin{bmatrix}
     1 & 0 & \cdots & 0 
   \end{bmatrix}\in \RR^{1 \times (r-1)}\\
   \Lambda&=\begin{bmatrix}
		0 & 1 & \cdots & 0 \\
		\vdots & \vdots & \ddots & \vdots \\
		0 & 0 & \cdots & 1 \\
		-\gamma_0 & -\gamma_1 & \cdots & -\gamma_{r-2}
	\end{bmatrix}\in \RR^{(r-1) \times (r-1)}\\
G&=\begin{bmatrix}
        0 & \cdots & 0 & 1 
      \end{bmatrix}^T \in \RR^{(r-1) \times 1}\\
C(w)&=\begin{bmatrix}
        C_{1}(w) & C_{2}(w) & \cdots & C_{r-1}(w)  
      \end{bmatrix} \in \RR^{1\times (r-1)}
\end{align*}
in which, $C_{1}(w)=c_1(w)-(c_r(w)+\gamma_{r-2})\gamma_0$
and, for $s=2,\ldots, r-1$, 
$C_{s}(w)=c_s(w)+\gamma_{s-2}-(c_r(w)+\gamma_{r-2})\gamma_{s-1}$, 
and the rest parameters are given by
\begin{align*}
		\tilde c(w,k_1)&=c_r(w)+\gamma_{r-2}-k_1 b(w)+\Psi T^{-1}_1N_1 \\
		\check{A}_3(w)&=-\frac{1}{b(w)}N_1 A_3(w) \\
		\check{C}_1(w)&=-\frac{1}{b(w)}N_1 C(w)\\
		\check{C}_2(w)&=\frac{1}{b(w)}(M_1N_1-(c_r(w)+\gamma_{r-2})N_1)\\
		\bar A_3(w,k_1)&=k_1A_3(w) \\
		\bar C_1(w,k_1)&=k_1 C(w) \\
		\bar c_1(w,k_1)&=k_1(\tilde c(w,k_1)-a+\Psi T^{-1}_1N_1) \\ 
		&\quad\ +\frac{1}{b(w)}\Psi T^{-1}_1 (aI_l-M_1-N_1\Psi T^{-1}_1)N_1 \\
		\bar C_2(w,k_1)&=\Psi T^{-1}_1 \left((a+k_1 b(w))I_l-M_1-N_1\Psi T^{-1}_1\right) \\
		\bar c_2(w,k_1)&= a -\Psi T^{-1}_1N_1+\Psi T^{-1}_2N_2+ k_1 b(w) \\
		\hat A_3(w,k_1)&=-\frac{1}{b_a}N_2 \bar A_3(w,k_1)\\
		\hat C_1(w,k_1)&=-\frac{1}{b_a}N_2\bar C_1(w,k_1)\\
		\hat C_2(w,k_1)&=-\frac{1}{b_a}N_2\bar c_1(w,k_1)\\
		\hat C_3(w,k_1)&=-\frac{1}{b_a}N_2\bar C_2(w,k_1)\\
		\hat C_4(w,k_1)&= \frac{1}{b_a}\left((M_2+N_2\Psi T^{-1}_2)N_2-N_2\bar c_2(w,k_1)\right).
\end{align*}

\subsection{Solvability of Problem \ref{Problem-Robust_output_regulation}}

\begin{Lemma}\label{Result_of_state_feedback}
Under Assumption \ref{Ass-nonnegative}, there exist a positive constant $k^{*}_1$ and a positive function $\phi_1(k_1)$ such that, for any $k_2>\phi_1(k_1)$ with $k_1>k^{*}_1$, 
the robust stabilization problem of the augmented system \eqref{eq-augmented_system_transformed} is solved by the following static state feedback control law:
	\begin{align}\label{eq-control_law_augmented_system}
		\bar u_1 &=-k_2\zeta_2     \notag\\
		\zeta_2 &=\bar x_1+k_1\zeta_1   \notag\\
		\zeta_1 &=e^{(r-1)}+\gamma_{r-2} e^{(r-2)}+\cdots+\gamma_1 e^{(1)}+\gamma_0 e
	\end{align}
where $\gamma_s, s=0,1,\dots,r-2$, are positive constants such that the polynomial $f(\lambda)$ defined in \eqref{eq-stable_polynomial} is stable.
\end{Lemma}

\begin{Proof}
    See Appendix B.
\end{Proof}

It is noted that the control law \eqref{eq-control_law_augmented_system} relies on the regulated error $e(t)$ and its derivatives $e^{(i)}(t), i=1,\dots,r-1$. 
Hence, the control law \eqref{eq-control_law_augmented_system} is not implementable
when the derivatives of the regulated error are not measurable. 
To address this issue, inspired by \cite{Chen-Huang-2015book}, 
we further design a dynamic output feedback control law based on a high-gain observer. The result is summarized as follows.

\begin{Lemma}\label{Result_of_output_feedback}
Under Assumption \ref{Ass-nonnegative}, 
in addition to the existence of $k_1^*>0$ and $\phi_1(k_1)>0$ 
in Lemma \ref{Result_of_state_feedback},  
there further exists a positive function $\phi_2(k_2)$ such that, for any $h>\mathrm{max}\{1, \phi_2(k_2)\}$ and $k_2>\phi_1(k_1)$ with $k_1>k^{*}_1$, the robust stabilization problem of the augmented system \eqref{eq-transformation-2} is solved by the following dynamic output feedback control law:
	\begin{align}\label{eq-output_control_law_augmented_system}
		\bar u_1 &=-k_2\hat \zeta_2     \notag\\
		\hat\zeta_2 &=\bar x_1+k_1\hat \zeta_1   \notag\\
		\hat\zeta_1 &=\varsigma_r+\gamma_{r-2} \varsigma_{r-1}+\cdots+\gamma_1 \varsigma_2+\gamma_0 \varsigma_1 \notag\\
		\dot\varsigma&=A_0(h)\varsigma+B_0(h)e                     
	\end{align}
where $\gamma_s, s=0,1,\dots,r-2$, 
are positive constants such that the polynomial $f(\lambda)$ defined in \eqref{eq-stable_polynomial} is stable,
$\varsigma=\mathrm{col}\left(\varsigma_1,\dots,\varsigma_r\right)\in\RR^r$,
and
	\begin{align}\label{eq-A0-B0-def}
		A_0(h)&=
		\begin{bmatrix}
			-h \delta_{r-1} & 1 & 0 & \cdots & 0 \\
			-h^2 \delta_{r-2} & 0 & 1 & \cdots & 0 \\
			\vdots & \vdots & \vdots & \ddots & \vdots \\
			-h^{r-1} \delta_1 & 0 & 0 & \cdots & 1 \\
			-h^r \delta_0 & 0 & 0 & \cdots & 0
		\end{bmatrix}\in\RR^{r\times r} \notag\\
		 B_0(h)&=
		\begin{bmatrix}
			h \delta_{r-1} \\
			h^2 \delta_{r-2} \\
			\vdots \\
			h^{r-1} \delta_1 \\
			h^r \delta_0
		\end{bmatrix}\in\RR^{r\times 1}
	\end{align}
in which, $\delta_s, s=0,1,\dots,r-1$, are positive constants such that the polynomial 
	\begin{align}\label{eq-def-g(s)}
		g(\lambda)=\lambda^{r}+\delta_{r-1}\lambda^{r-1}+\dots+\delta_1 \lambda+\delta_0
	\end{align} is stable.
\end{Lemma}

\begin{Proof}
	See Appendix C.
\end{Proof}

Combining Lemma \ref{Result_of_state_feedback}
and Lemma \ref{Result_of_output_feedback} leads to the solvability of Problem \ref{Problem-Robust_output_regulation} as stated below.

\begin{Theorem}\label{Thorem-single-actuator}
	Under Assumption \ref{Ass-nonnegative}, there exist a positive constant $k^{*}_1$ 
and positive functions $\phi_1(k_1)$ and $\phi_2(k_2)$ such that, for any
	\begin{align}\label{ieq-h-k1-k2-theorem1}
	k_1>k^{*}_1,	\quad  k_2>\phi_1(k_1),   \quad h>\mathrm{max}\{1,\phi_2(k_2)\}
	\end{align}
Problem \ref{Problem-Robust_output_regulation} is solved by the following dynamic output feedback control law:
	\begin{align}\label{eq-control-law-single-actuator}
		u_1 &=\Psi T^{-1}_2\eta_2-k_2(x_1-\Psi T^{-1}_1\eta_1+k_1\hat \zeta_1)     \notag\\
		\hat\zeta_1 &=\varsigma_r+\gamma_{r-2} \varsigma_{r-1}+\cdots+\gamma_1 \varsigma_2+\gamma_0 \varsigma_1 \notag\\
		\dot\eta_1&=M_1\eta_1+N_1x_1  \notag\\
		\dot\eta_2&=M_2\eta_2+N_2u_1   \notag\\
		\dot\varsigma&=A_0(h)\varsigma+B_0(h)e.                
	\end{align}
\end{Theorem}

\begin{Remark}
It is interesting to compare our result with 
the existing results on linear robust output regulation without considering the actuator
and linear output regulation with a single ``virtual'' actuator. 
First, unlike the dynamic output feedback control law proposed in \cite{Chen-Huang-2015book} for linear robust output regulation without considering the actuator,  our control law~\eqref{eq-control-law-single-actuator} requires the design 
of an additional internal model, i.e., the $\eta_1$-subsystem, 
and an extra design parameter $k_1$. 
Second, different from the feedforward control approach adopted in \cite{Wei_Li_Wang-2023TCYB} for linear output regulation with a single ``virtual'' actuator, 
our control law is based on the internal model approach and the high-gain observer technique.
This enables our control law to handle uncertain linear plants 
with arbitrarily large uncertainties within a prescribed compact set.
\end{Remark}

\begin{Remark}
	The design details for $k_1^*$, $\phi_1(\cdot)$, and $\phi_2(\cdot)$ can be found in Appendices B and C. These parameters depend heavily on the parameter uncertain vector set $\mathbb{W}$ 
	and the bounds of the solutions to the closed-loop parameterized Lyapunov equations, 
	which may lead to conservativeness in their general expressions. 
	Nevertheless, once the plant and the uncertain parameter vector set $\mathbb{W}$ are specified, these parameters can be explicitly determined. 
\end{Remark}

\section{Robust output regulation with cooperative control of parallel actuators}\label{Section-MR}

In this section, we study the main problem of this paper,
namely, Problem \ref{Problem-Robust_output_regulation_multiple_actuators}. 
Compared with Problem \ref{Problem-Robust_output_regulation}, 
Problem \ref{Problem-Robust_output_regulation_multiple_actuators}
requires an additional Property \eqref{Property-three-multiple}, i.e., the input sharing property. 
To fulfill this requirement, 
similar to what has been shown in \cite{Lim_Oh-2024Auto} and \cite{Xu_Su_Liu-2025Auto}, 
coupling terms need to be incorporated into the controllers of these actuators. 
Specifically,  we propose the following distributed internal model:
\begin{subequations}\label{Internal-Model-i}
  \begin{align} \dot\eta_{1i}&=M_1\eta_{1i}+N_1x_i+\sigma_1\sum_{j\in\mathcal{N}_i}a_{ij}(\eta_{1j}-\eta_{1i})   \\	\dot\eta_{2i}&=M_2\eta_{2i}+N_2u_i+\sigma_2\sum_{j\in\mathcal{N}_i}a_{ij}(\eta_{2j}-\eta_{2i})
\end{align} 
\end{subequations}
where, for $i=1,\dots, N$, $\eta_{1i}\in\RR^{l}$ and $\eta_{2i}\in\RR^{l}$ are the states, 
$M_1, M_2, N_1$, and $N_2$ are constant matrices, same as those defined
in \eqref{Internal_Model}, and $\sigma_1$ and $\sigma_2$ are positive constants to be determined. Moreover, the elements $a_{ij}$ are entries of the weighted adjacency matrix $\mathcal{A}=[a_{ij}]_{i,j=1}^{N} \in \RR^{N\times N}$ of the graph $\mathcal{G}$.
In particular, $a_{ii}=0$ and $a_{ij}>0$ if and only if $(j,i)\in \mathcal{E}, i,j \in \mathcal{V}$.

Further, we design high-gain observers as follows:
\begin{align}\label{eq-distributed-observer}
	\dot{\hat\varsigma}_i=A_0(\bar h)\hat\varsigma_i+B_0(\bar h)e   
\end{align}
where, for $i=1,\dots, N$, $\hat\varsigma_i\in\RR^{r}$ is the state, 
$A_0(\bar h)$ and $B_0(\bar h)$ are two matrices in the same form of \eqref{eq-A0-B0-def}.

Then, on the basis of the control law \eqref{eq-control-law-single-actuator} for a single actuator, we propose the following distributed dynamic output feedback control law:
\begin{align}\label{eq-distributed-output-feedback-control-law}
	u_i &=\Psi T^{-1}_2\eta_{2i}-\bar k_2(x_i-\Psi T^{-1}_1\eta_{1i}+\bar k_1\hat \zeta_{1i})     \notag\\
	\hat\zeta_{1i} &=\hat\varsigma_{ri}+\gamma_{r-2} \hat\varsigma_{(r-1)i}+\cdots+\gamma_1\hat\varsigma_{2i}+\gamma_0 \hat\varsigma_{1i} \notag\\
	\dot\eta_{1i}&=M_1\eta_{1i}+N_1x_i+\sigma_1\sum_{j\in\mathcal{N}_i}a_{ij}(\eta_{1j}-\eta_{1i})   \notag\\
	\dot\eta_{2i}&=M_2\eta_{2i}+N_2u_i+\sigma_2\sum_{j\in\mathcal{N}_i}a_{ij}(\eta_{2j}-\eta_{2i})     \notag\\
	\dot{\hat\varsigma}_i&=A_0(\bar h)\hat\varsigma_i+B_0(\bar h)e, \quad i=1,\dots, N
\end{align}
where $\gamma_s, s=0,1,\dots,r-2$ are defined in \eqref{eq-control_law_augmented_system}, $\hat\varsigma_i=\mathrm{col}\left(\hat\varsigma_{1i}, \dots, \hat\varsigma_{ri}\right)$, 
and $\bar k_1$, $\bar k_2$, $\bar h$, as well as $\sigma_1$ and $\sigma_2$, are all positive constants to be determined.

Next, define the sum state $ x_{a} \in \RR^{N}$ 
and the sum input $u \in \RR^{m}$ of the $N$ actuators in \eqref{eq-multiple-actuators}, 
and the sum states $\eta_{1}\in\RR^{l}$, $\eta_{2}\in\RR^{l}$ of the distributed internal model \eqref{Internal-Model-i} as follows:
\begin{align}\label{eq-sum-state-input}
	x_{a}&=\sum_{i=1}^N x_i,   &  u &= \sum_{i=1}^N u_i \notag \\
	\eta_1&=\sum_{i=1}^N \eta_{1i}, &  \eta_2 &= \sum_{i=1}^N \eta_{2i}. 
\end{align}
The so-called sum dynamics of the $N$ actuators in \eqref{eq-multiple-actuators} 
and those of the distributed internal model \eqref{Internal-Model-i} are described by
  \begin{align}\label{eq-actuator-sum}
	\dot{x}_{a} &=ax_{a}+b_au  \notag\\
	u_p&=x_a 
\end{align}
and 
\begin{subequations}\label{eq-Internal_Model-sum}
\begin{align}
	\dot\eta_1&=M_1\eta_1+N_1x_a-\sigma_1\left(\mathbf{1}^{T}_{N}\LL\otimes I_l\right)\hat\eta_1  \\
	\dot\eta_2&=M_2\eta_2+N_2u-\sigma_2 \left(\mathbf{1}^{T}_{N}\LL \otimes I_l\right)\hat\eta_2 
\end{align}
\end{subequations}
where $\hat\eta_1=\mathrm{col}\left(\eta_{11}, \dots, \eta_{1N}\right)$, $\hat\eta_2=\mathrm{col}\left(\eta_{21}, \dots, \eta_{2N}\right)$, and $\mathcal{L}=[l_{ij}]_{i,j=1}^{N} \in \RR^{N \times N}$ 
is the Laplacian matrix of the graph $\mathcal{G}$,
defined from its weighted adjacency matrix $\mathcal{A}$
as $l_{ii}=\sum_{j=1}^{N} a_{ij}$ and $l_{ij}=-a_{ij}, i\ne j$.
Since the graph $\G$ is assumed to be undirected, $\LL$ is symmetric. 
Thus, $\LL\mathbf{1}_{N}=0$ implies that 
\begin{align}\label{eq-1-kronecker-L}
	 \mathbf{1}^{T}_{N}\LL=\zero.
\end{align}

Attaching \eqref{eq-actuator-sum} and \eqref{eq-Internal_Model-sum} with \eqref{eq-1-kronecker-L}
to the plant \eqref{eq-plant} gives the following augmented system:
\begin{align}\label{eq-augmented-system-multiple}
	\dot z & =A_1(w) z+A_2(w) \xi_1+E_0(w) v \notag\\
	\dot{\xi}_s & =\xi_{(s+1)}, \quad s=1, \ldots, r-1 \notag\\
	\dot{\xi}_r & =A_3(w) z+\sum_{s=1}^r c_s(w) \xi_s+E_r(w) v+b(w) x_a \notag\\
	\dot{x}_{a} &=ax_{a}+b_au \notag \\
	\dot\eta_1&=M_1\eta_1+N_1x_a  \notag \\
	\dot\eta_2&=M_2\eta_2+N_2u   \notag\\
	e & =\xi_1-F(w)v.
\end{align}

Now, define the average state $\bar \varsigma\in\RR^{r}$ of the observers in \eqref{eq-distributed-observer} as follows: 
\begin{align}\label{eq-average-varsigma}
	\bar\varsigma=\frac{1}{N}\sum_{i=1}^{N}\hat\varsigma_i.
\end{align}
Then, the so-called average dynamics of the observers in \eqref{eq-distributed-observer} can be expressed as
\begin{align}\label{eq-dot-average-varsigma}
	\dot{\bar\varsigma}=A_0(\bar h)\bar\varsigma+B_0(\bar h)e.
\end{align}

Finally, denote $\bar\zeta_1=\frac{1}{N}\sum_{i=1}^{N}\hat\zeta_{1i}$ 
and $\bar \varsigma=\mathrm{col}\left(\bar\varsigma_1,\dots,\bar\varsigma_r\right)$. 
From \eqref{eq-distributed-output-feedback-control-law}, \eqref{eq-sum-state-input}, \eqref{eq-Internal_Model-sum}, \eqref{eq-1-kronecker-L}, \eqref{eq-average-varsigma}, 
and \eqref{eq-dot-average-varsigma}, the sum input $u$ to the augmented system \eqref{eq-augmented-system-multiple} 
is generated by
\begin{align}\label{eq-distributed-output-feedback-control-law-sum}
	u &=\Psi T^{-1}_2\eta_2-\bar k_2(x_a-\Psi T^{-1}_1\eta_1+N\bar k_1\bar \zeta_1)    \notag\\
	\bar\zeta_1 &=\bar\varsigma_r+\gamma_{r-2} \bar\varsigma_{r-1}+\cdots+\gamma_1\bar\varsigma_2+\gamma_0\bar \varsigma_1 \notag\\
	\dot\eta_1&=M_1\eta_1+N_1x_a\notag \\
	\dot\eta_2&=M_2\eta_2+N_2u \notag\\
	\dot{\bar\varsigma}&=A_0(\bar h)\bar\varsigma+B_0(\bar h)e.
\end{align}

\subsection{Robust Output Regulation}\label{subsection-ROR}
It is noted that the sum input  $u$ to the augmented system \eqref{eq-augmented-system-multiple} as generated by \eqref{eq-distributed-output-feedback-control-law-sum} does not explicitly contain the coupling terms.
In other words, Properties \eqref{Property-one-multiple} and \eqref{Property-two-multiple} on the closed-loop system are independent of the design of the parameters $\sigma_1$ and $\sigma_2$ for the coupling terms in \eqref{eq-distributed-output-feedback-control-law}. 
Thus, in this subsection, we first present our design for 
$\bar{k}_1$, $\bar{k}_2$, and $\bar{h}$ to ensure the satisfaction of Properties \eqref{Property-one-multiple} and \eqref{Property-two-multiple} of Problem \ref{Problem-Robust_output_regulation_multiple_actuators}.

\begin{Lemma}\label{Lemma-robust-output-regulation-multiple-actuators}
	Under Assumption \ref{Ass-nonnegative}, for any positive constant $k^{*}_1$ and positive functions $\phi_1(\cdot)$ and $\phi_2(\cdot)$ asserted by Theorem~\ref{Thorem-single-actuator}, choose $\bar k_1, \bar k_2$, and $\bar h$ such that
	\begin{align}\label{eq-bark1-k2-h-def}
		 N\bar k_1> k^{*}_1, \quad  \bar k_2>\phi_1(N\bar k_1), \quad \bar h>\mathrm{max}\{1,\phi_2(\bar k_2)\}.
	\end{align}
	Then, Properties \eqref{Property-one-multiple} and \eqref{Property-two-multiple} of Problem \ref{Problem-Robust_output_regulation_multiple_actuators} are achieved by the distributed dynamic output feedback control law \eqref{eq-distributed-output-feedback-control-law}.
\end{Lemma}

\begin{Proof}
First, we note that  
the sum input $u$ of the distributed dynamic output feedback control law \eqref{eq-distributed-output-feedback-control-law} for multiple actuators 
is generated by \eqref{eq-distributed-output-feedback-control-law-sum}.
By comparing \eqref{eq-control-law-single-actuator} and \eqref{eq-distributed-output-feedback-control-law-sum}, 
we know that \eqref{eq-distributed-output-feedback-control-law-sum} is in the same form as the dynamic output feedback control law \eqref{eq-control-law-single-actuator} for a single actuator. In particular, one only needs to replace 
$u, x_a, N\bar k_1, \bar k_2, \bar\zeta_1, \bar h$, and $\bar\varsigma$ in \eqref{eq-distributed-output-feedback-control-law-sum}
with $u_1, x_1, k_1, k_2, \hat\zeta_1, h$, and $\varsigma$ in \eqref{eq-control-law-single-actuator}, respectively. 

Furthermore, we note that the augmented system \eqref{eq-augmented-system-multiple} with multiple actuators maintains the same form as the augmented system \eqref{eq-augmented-system-single} for a single actuator, by replacing $x_a$ and $u$ in
\eqref{eq-augmented-system-multiple} 
with $x_1$ and $u_1$ in \eqref{eq-augmented-system-single}, respectively.

Therefore, using Theorem \ref{Thorem-single-actuator}, the design of $\bar k_1$, $\bar k_2$, and $\bar h$ in \eqref{eq-bark1-k2-h-def}
for the distributed dynamic output feedback control law~\eqref{eq-distributed-output-feedback-control-law}
achieves Properties \eqref{Property-one-multiple} and \eqref{Property-two-multiple} of Problem \ref{Problem-Robust_output_regulation_multiple_actuators}.
\end{Proof}

\begin{Remark}\label{Remark-design-bark1-k2-h}
A general design of $\bar k_1, \bar k_2$, and $\bar h$ as in \eqref{eq-bark1-k2-h-def} 
seems to be dependent on the number of actuators $N$. 
Nevertheless, this dependency can be significantly relaxed by
a simple design as follows:
\begin{align*} 
	\bar{k}_1 = \frac{k_1}{N}, \quad \bar{k}_2 = k_2, \quad \bar{h} = h
\end{align*} 
where $k_1$, $k_2$, and $h$ are any parameters that satisfy \eqref{ieq-h-k1-k2-theorem1}.
This implies that the parameters $k_1$, $k_2$, and $h$ designed in the dynamic output feedback control law \eqref{eq-control-law-single-actuator} for a single actuator can be directly used in the distributed dynamic output feedback control law \eqref{eq-distributed-output-feedback-control-law} for multiple actuators to achieve Properties~\eqref{Property-one-multiple} and \eqref{Property-two-multiple} of Problem \ref{Problem-Robust_output_regulation_multiple_actuators}. 
Interestingly, only the parameter $\bar k_1$ depends on the number of actuators $N$
in a very simple way.
\end{Remark}

As noted in the beginning of this subsection, 
Properties~\eqref{Property-one-multiple} and \eqref{Property-two-multiple} of 
Problem~\ref{Problem-Robust_output_regulation_multiple_actuators} hinge only on the design of $\bar k_1, \bar k_2$, and $\bar h$, and are not affected by the design of the parameters $\sigma_1$ and $\sigma_2$. 
In contrast, the design of $\sigma_1$ and $\sigma_2$ is for achieving Property~\eqref{Property-three-multiple} of Problem~\ref{Problem-Robust_output_regulation_multiple_actuators}
 will be affected by the design of $\bar k_2$. The details are provided in the next subsection.

\subsection{Input Sharing Property}

Since the input sharing property is essentially the synchronization property of the actuators,
before presenting the design of the parameters $\sigma_1$ and $\sigma_2$ 
for achieving Property \eqref{Property-three-multiple} of Problem \ref{Problem-Robust_output_regulation_multiple_actuators}, 
let us first deduce the closed-loop system of the $N$ actuators.

Applying the distributed dynamic output feedback control law \eqref{eq-distributed-output-feedback-control-law}
to the $N$ actuators in \eqref{eq-multiple-actuators} 
yields the following closed-loop system:
\begin{align}\label{eq-closed-loop-actuator-system}
	\dot x_i &=a x_i+b_a\left(\Psi T^{-1}_2\eta_{2i}-\bar k_2(x_i-\Psi T^{-1}_1\eta_{1i}+\bar k_1\hat \zeta_{1i})\right) \notag\\
	\hat\zeta_{1i} &=\hat\varsigma_{ri}+\gamma_{r-2} \hat\varsigma_{(r-1)i}+\cdots+\gamma_1 \hat\varsigma_{2i}+\gamma_0 \hat\varsigma_{1i} \notag\\
	\dot\eta_{1i}&=M_1\eta_{1i}+N_1x_i+\sigma_1\sum_{j\in\mathcal{N}_i}a_{ij}(\eta_{1j}-\eta_{1i})   \notag\\
	\dot\eta_{2i}&=M_2\eta_{2i}+N_2\left(\Psi T^{-1}_2\eta_{2i}-\bar k_2(x_i-\Psi T^{-1}_1\eta_{1i}+\bar k_1\hat \zeta_{1i})\right)\notag\\
	&\quad +\sigma_2\sum_{j\in\mathcal{N}_i}a_{ij}(\eta_{2j}-\eta_{2i})    \notag \\
	\dot{\hat\varsigma}_i&=A_0(\bar h)\hat\varsigma_i+B_0(\bar h)e, \quad i=1,\dots, N.   
\end{align}
Let $\hat x_i=\mathrm{col}\left(x_i, \eta_{1i}, \eta_{2i}\right) \in \RR^{2l+1}$ and
\begin{align}\label{eq-hat-xi-J-Gamma-def}
	J&=\mathrm{block~diag}\left\{0,\sigma_1I_l, \sigma_2I_l \right\} \notag\\
	\Upsilon&=\begin{bmatrix}
		-b_a\bar k_1\bar k_2\Gamma_1  \\ 
		 \zero  \\
		-\bar k_1\bar k_2N_2\Gamma_1
	\end{bmatrix}
\end{align}
in which, $\Gamma_1=\begin{bmatrix}
                \gamma_0 & \gamma_1 & \cdots & \gamma_{r-2} & 1 
              \end{bmatrix}$.
Then, a compact form of the closed-loop system \eqref{eq-closed-loop-actuator-system} 
can be obtained as follows:
\begin{align}\label{eq-closed-loop-actuator-system-i}
	\dot {\hat x}_i
	&=A\hat x_i+ J \sum_{j\in\mathcal{N}_i}a_{ij}(\hat x_j-\hat x_i)+\Upsilon \hat\varsigma_i \notag\\
\dot{\hat\varsigma}_i&=A_0(\bar h)\hat\varsigma_i+B_0(\bar h)e, \quad i=1,\dots, N   
\end{align}
where
\begin{align}\label{eq-A-def}
	A=\begin{bmatrix}
		a-b_a\bar k_2 & b_a\bar k_2\Psi T^{-1}_1  & b_a\Psi T^{-1}_2 \\
		N_1      &  M_1       &   \zero  \\
		-N_2\bar k_2  &  \bar k_2N_2\Psi T^{-1}_1  &  M_2+N_2\Psi T^{-1}_2 
	\end{bmatrix}.
\end{align}

Now, we are ready to present our design of $\sigma_1$ and $\sigma_2$.

\begin{Lemma}\label{lemma-sigma-design}
Under Assumptions \ref{Ass-nonnegative} and \ref{Ass-connected-graph}, 
given any positive constants $\bar k_1$, $\bar k_2> \frac{a}{b_a} $, and $\bar h$, 
there exists a positive constant $\sigma^{*}(\bar k_2)$ such that, for any
	\begin{align}\label{ieq-sigma1-2-3-design}
		\min\{\sigma_1, \sigma_2\}>\sigma^{*}(\bar k_2) 
	\end{align}
Property~\eqref{Property-three-multiple} of Problem \ref{Problem-Robust_output_regulation_multiple_actuators} is achieved by the distributed dynamic output feedback control law \eqref{eq-distributed-output-feedback-control-law}.
\end{Lemma}

\begin{Proof}
Denote $\hat x=\mathrm{col}\left(\hat x_1,\dots,\hat x_N \right)\in \RR^{N(2l+1)}$ and $\hat\varsigma=\mathrm{col}(\hat\varsigma_1,\dots,\hat\varsigma_N)\in \RR^{Nr}$. 
Then, the closed-loop system \eqref{eq-closed-loop-actuator-system-i} 
can be put into the following more compact form: 
\begin{align}\label{eq-dynamic-hat-X}
	\begin{bmatrix}
		\dot {\hat x} \\ \dot {\hat \varsigma}
	\end{bmatrix}
	&=\begin{bmatrix}
		I_N\otimes A-\LL\otimes J & I_N\otimes\Upsilon  \\
		\zero  &  I_N\otimes A_0(\bar h)
	\end{bmatrix}
	\begin{bmatrix}
		\hat x \\  \hat\varsigma
	\end{bmatrix}\notag\\
&\quad +
	\begin{bmatrix}
		\zero \\ \mathbf{1}_{N}\otimes B_0(\bar h)e
	\end{bmatrix}.
\end{align}
Like in \cite{Xu_Su_Liu-2025Auto}, we define 
$\LL_{0}=I_{N}-\frac{\mathbf{1}_{N}\mathbf{1}^{T}_{N}}{N}$,
which clearly satisfies $\LL_{0}\mathbf{1}_{N}=0$ and  $\mathbf{1}_{N}^{T}\LL_{0}=\zero$.
Moreover, it can be verified that $\LL_{0}\LL=\LL \LL_{0}=\LL$.
	Let 
	\begin{align}\label{eq-hat-X-def}
		\hat X= 
\begin{bmatrix}
  \LL_0 \otimes I_{2l+1} & \zero \\
  \zero & \LL_0 \otimes I_{r}
\end{bmatrix}
	    \begin{bmatrix}
			\hat x \\  \hat\varsigma
		\end{bmatrix}.
	\end{align}
It follows from \eqref{eq-dynamic-hat-X} that 
\begin{align}\label{eq-der-L0-otimes-z}
	\dot{\hat X}
	&=\begin{bmatrix}
		\LL_0 \otimes A-\LL \LL_0 \otimes J & \LL_0 \otimes\Upsilon  \\
		\zero  &  \LL_0 \otimes A_0(\bar h)
	\end{bmatrix}\begin{bmatrix}
		\hat x \\  \hat\varsigma
	\end{bmatrix} \notag\\
	&\ \quad
	+
	\begin{bmatrix}
		\zero \\ \LL_0\mathbf{1}_{N}\otimes B_0(\bar h)e
	\end{bmatrix}  \notag\\
	&= \begin{bmatrix}
		I_N\otimes A-\LL\otimes J & I_N\otimes\Upsilon  \\
		\zero  &  I_N\otimes A_0(\bar h)
	\end{bmatrix}
	\hat X. 
\end{align}

Under Assumption \ref{Ass-connected-graph}, the Laplacian matrix $\LL$ has a simple eigenvalue $\lambda_{1}=0$ and $(N-1)$ positive eigenvalues $\lambda_{i}$, $i=2,\ldots,N$.
Thus, we can find a nonsingular matrix $\Gamma_2 \in\RR^{N\times N}$ of the following form:
	\begin{align*}
		\Gamma_2=\begin{bmatrix} \mathbf{1}_{N} & V \end{bmatrix}, \qquad  \Gamma^{-1}_2=\begin{bmatrix} \frac{1}{N}\mathbf{1}^{T}_{N} \\ V^{\dagger} \end{bmatrix}
	\end{align*}
	in which, $V\in\RR^{N\times (N-1)}$ and $V^{\dagger}\in\RR^{(N-1)\times N}$, 
	such that
	\begin{align}\label{eq-T--1-L-T}
		\Gamma^{-1}_2\LL\Gamma_2=\mathrm{diag}\{0, \lambda_{2}, \dots, \lambda_{N}\}.
	\end{align}
Let 
	\begin{align}\label{eq-bf-hat-X-def}
		\mathbf{\hat X } =
\begin{bmatrix}
  \Gamma^{-1}_2\otimes I_{2l+1} & \zero \\
  \zero & \Gamma^{-1}_2\otimes I_{r}
\end{bmatrix} \hat X .
	\end{align} 
It further follows from \eqref{eq-der-L0-otimes-z} and \eqref{eq-T--1-L-T} that the time derivative of $\mathbf{\hat X}$ satisfies \eqref{eq-der-hatX}.
	\begin{figure*}
			\hrulefill
		\begin{align}\label{eq-der-hatX}
			\mathbf{\dot{\hat{X}}}
			&=\begin{bmatrix}
				I_N\otimes A-\Gamma^{-1}_2\LL\Gamma_2\otimes J & I_N\otimes\Upsilon  \notag\\
				\zero  &  I_N\otimes A_0(\bar h)
			\end{bmatrix} 
			\mathbf{\hat X} \notag\\
			&=\begin{bmatrix}
				\begin{array}{cc|cc}
					A & \zero & \Upsilon & \zero  \\ 
					\zero & I_{N-1} \otimes A -\mathrm{diag}\{\lambda_{2}, \dots, \lambda_{N}\} \otimes J & \zero & I_{N-1}\otimes\Upsilon \\ 
					\hline 
					\zero & \zero & A_0(\bar   h) & \zero   \\
					\zero & \zero & \zero     &  I_{N-1} \otimes A_0(\bar h) 
				\end{array}
			\end{bmatrix}\mathbf{\hat X}.
		\end{align}
		\hrulefill
	\end{figure*}

	Next, we partition 
\begin{equation*}
  \mathbf{\hat X}  = \mathrm{col}\left(\mathbf{\hat{X}}_1, \mathbf{\hat{X}}_2, 
  \mathbf{\hat{X}}_3, \mathbf{\hat{X}}_4\right)\in\RR^{N(2l+1+r)}
\end{equation*}
with $\mathbf{\hat{X}}_1\in\RR^{2l+1}$, 
$\mathbf{\hat{X}}_2\in\RR^{(N-1)(2l+1)}$, 
$\mathbf{\hat{X}}_3\in\RR^{r}$, and $\mathbf{\hat{X}}_4\in\RR^{(N-1)r}$. 
From the definitions of $\hat X$ and $\mathbf{\hat X}$ in \eqref{eq-hat-X-def} and \eqref{eq-bf-hat-X-def}, respectively, we have
	\begin{align}\label{eq-X1-X3=0}
		\mathbf{\hat X}_1&=\left(\frac{1}{N}\mathbf{1}^{T}_{N}\otimes I_{2l+1}\right)(\LL_{0}\otimes I_{2l+1})\hat x \equiv 0   \notag\\
		\mathbf{\hat X}_3&=\left(\frac{1}{N}\mathbf{1}^{T}_{N}\otimes I_l\right)(\LL_{0}\otimes I_l)\hat\varsigma \equiv 0.
	\end{align}
In addition, from \eqref{eq-der-hatX}, we know that 
\begin{align}\label{eq-subsystem-hat-X2-X4}
	{\mathbf{\dot{\hat{X}}}}_2 &=\left(I_{N-1} \otimes A -\mathrm{diag}\{\lambda_{2}, \dots, \lambda_{N}\} \otimes J\right) \mathbf{\hat X_2} \notag\\
	 &\quad + (I_{N-1}\otimes\Upsilon)\mathbf{\hat X_4} \notag\\
	{\mathbf{\dot{\hat{X}}}}_4&=\left(I_{N-1} \otimes A_0(\bar h) \right) \mathbf{\hat X_4}.
\end{align}

Recall the definition of $A_0(\cdot)$ in \eqref{eq-A0-B0-def} and \eqref{eq-def-g(s)}. For any $\bar h>0$, it is clear that $A_0(\bar h)$ is Hurwitz and so is the following block diagonal matrix:
\begin{align}\label{eq-system-matrix-A0-sigma3}
	 I_{N-1} \otimes A_0(\bar h).
\end{align}

Further, from the definitions of $J$ and $A$ in \eqref{eq-hat-xi-J-Gamma-def} and \eqref{eq-A-def}, respectively, for $i=2,\dots, N$, we have
\begin{align}\label{eq-A-lambda-J}
	A-\lambda_{i}J=
	\begin{bmatrix}
		A_{11}(\bar k_2)  &  A_{12}(\bar k_2) \\
		A_{21}(\bar k_2)  &  A_{22}(\bar k_2)+\sigma B(\lambda_{i})
	\end{bmatrix}
\end{align}
where 
\begin{align*}
	A_{11}(\bar k_2)&=a-b_a\bar k_2, \quad\, A_{12}(\bar k_2)=\begin{bmatrix}
		b_a\bar k_2\Psi T^{-1}_1  & b_a\Psi T^{-1}_2
	\end{bmatrix} \notag\\
	A_{21}(\bar k_2)&=\begin{bmatrix}
		N_1   \\  -N_2\bar k_2
	\end{bmatrix}, \ \quad B(\lambda_i)=-\lambda_iI_{2l} \notag\\
	A_{22}(\bar k_2)&=\begin{bmatrix}
		M_1   &   \zero   \\
		\bar k_2N_2\Psi T^{-1}_1  &  M_2+N_2\Psi T^{-1}_2
	\end{bmatrix}  \notag\\
	\sigma&=\mathrm{block~diag}\{\sigma_1I_l, \sigma_2I_l\}. 
\end{align*}
It is clear that $\Vert A_{12}(\bar k_2) \Vert $, $\Vert A_{21}(\bar k_2) \Vert $, and $\Vert A_{22}(\bar k_2) \Vert $ are all bounded by some positive functions of $\bar k_2$. 
Since $\lambda_i>0$ for all $ i=2\dots,N$, $B(\lambda_i)$ is Hurwitz. 
Also, letting $\bar k_2>\frac{a}{b_a}$ ensures that $A_{11}(\bar k_2)$ is Hurwitz.

It is noted the matrix \eqref{eq-A-lambda-J} takes the form of \eqref{eq-matrix_lemma} 
with $M_1(w,\mu_1)=A_{11}(\bar k_2)$, $M_2(w)=B(\lambda_{i})$, $N_1(w,\mu_1)=A_{12}(\bar k_2)$, $N_2(w,\mu_1)=A_{21}(\bar k_2)$, $N_3(w,\mu_1)=A_{22}(\bar k_2)$, $\mu_1=\bar k_2$, $\mu_2=\sigma$, and
$w\in \{\lambda_2,\ldots, \lambda_N\}$.
Thus, by Lemma~\ref{Lemma_asymptotically_stable_system} in Appendix \ref{Appendix-lemma}, 
given any positive constant $\bar k_2>\frac{a}{b_a}$,
there exists a positive function $\sigma^{*}(\cdot)$ of the form $\phi(\cdot)$ defined in \eqref{eq-phi-function-def} such that, for any $\min\{\sigma_1, \sigma_2\}>\sigma^*(\bar k_2)$, the matrix $A-\lambda_iJ$ is Hurwitz for all $i=2,\ldots,N$. 
Therefore, the following block diagonal matrix is Hurwitz:
\begin{align}\label{eq-system-matrix-A-J}
	I_{N-1} \otimes A -\mathrm{diag}\{\lambda_{2}, \dots, \lambda_{N}\} \otimes J.
\end{align}

Now, since system \eqref{eq-subsystem-hat-X2-X4} is a cascaded linear system of 
the $\mathbf{\hat X}_2$- and $\mathbf{\hat X}_4$-subsystems, 
whose system matrices in \eqref{eq-system-matrix-A0-sigma3} and \eqref{eq-system-matrix-A-J} are both Hurwitz, 
system \eqref{eq-subsystem-hat-X2-X4} is asymptotically stable. 
By further noting \eqref{eq-X1-X3=0}, we have
$\lim_{t \to \infty} \mathbf{\hat X}(t)=0$,
which, together with \eqref{eq-hat-X-def} and \eqref{eq-bf-hat-X-def}, yields
\begin{align}\label{eq-limit-hat-X}
	\lim_{t \to \infty} \hat X(t)=\lim_{t\to\infty}\begin{bmatrix}
		(\LL_{0}\otimes I_{2l+1})\hat x(t) \\  (\LL_{0}\otimes I_{r})\hat\varsigma(t)
	\end{bmatrix}=0.
\end{align}
Finally, since the null space of the matrix $\LL_0\otimes I_{2l+1}$ is spanned by the columns of the matrix $\mathbf{1}_{N}\otimes I_{2l+1}$, \eqref{eq-limit-hat-X} implies that 
\begin{align}\label{eq-limit-har-xi-xj}
	\lim_{t \to \infty}(\hat x_{i}(t)-\hat x_{j}(t))=0, \quad \forall\, i,j=1,\dots, N.
\end{align}
Recall that the definition of $\hat x_{i}$ is $\hat x_i=\mathrm{col}\left(x_i, \eta_{1i}, \eta_{2i}\right)$
and $y_i=x_i$, $i=1,\dots, N$. 
It follows from \eqref{eq-limit-har-xi-xj} that 
\begin{align*}
	\lim_{t \to \infty}(y_{i}(t)-y_{j}(t))=0, \quad \forall\, i,j=1,\dots, N
\end{align*}
and hence, Property \eqref{Property-three-multiple} of Problem \ref{Problem-Robust_output_regulation_multiple_actuators} is satisfied.
\end{Proof}

\subsection{Solvability of Problem \ref{Problem-Robust_output_regulation_multiple_actuators}}

Based on Lemmas \ref{Lemma-robust-output-regulation-multiple-actuators} and \ref{lemma-sigma-design}, 
we establish the solvability of Problem \ref{Problem-Robust_output_regulation_multiple_actuators} 
as follows.

\begin{Theorem}\label{Theorem-multiple-actuator}
	Under Assumptions \ref{Ass-nonnegative} and \ref{Ass-connected-graph}, for any $k^*_1$ 
and positive functions $\phi_1(\cdot)$ and $\phi_2(\cdot)$ asserted by Theorem \ref{Thorem-single-actuator},
	there exists another positive function $\sigma^{*}(\cdot)$
such that, for any
\begin{align}\label{ieq-bark1-k2-h-sigma123}
	&N\bar k_1>k^*_1, \quad  \bar k_2>\mathrm{max}\left\{\phi_1(N\bar k_1),\frac{a}{b_a}\right\} \notag\\
	& \bar h>\mathrm{max}\{1,\phi_2(\bar k_2)\}, \quad  \mathrm{min}\{\sigma_1, \sigma_2\}>\sigma^{*}(\bar k_2)  
\end{align}
Problem \ref{Problem-Robust_output_regulation_multiple_actuators} is solved by the distributed dynamic output feedback control law \eqref{eq-distributed-output-feedback-control-law}.
\end{Theorem}

\begin{Proof}
	From Lemma \ref{Lemma-robust-output-regulation-multiple-actuators}, we know that Properties \eqref{Property-one-multiple} and \eqref{Property-two-multiple} of Problem \ref{Problem-Robust_output_regulation_multiple_actuators} are guaranteed when $\bar k_1$, $\bar k_2$, and $\bar h$ are designed according to \eqref{eq-bark1-k2-h-def},
which is repeated as follows:
\begin{align*}
	N\bar k_1> k^{*}_1,	\quad  \bar k_2>\phi_1(N\bar k_1), \quad   \bar h>\mathrm{max}\{1, \phi_2(\bar k_2)\}.
\end{align*}
    Building on Lemma \ref{Lemma-robust-output-regulation-multiple-actuators}, 
    it further follows from Lemma \ref{lemma-sigma-design} that Property \eqref{Property-three-multiple} of Problem \ref{Problem-Robust_output_regulation_multiple_actuators} is achieved when 
    $\bar k_2> \frac{a}{b_a}$ and $\sigma_1$ and $\sigma_2$ are designed
     according to \eqref{ieq-sigma1-2-3-design} as in the following:
    \begin{align*}
    	\min\{\sigma_1, \sigma_2\}>\sigma^{*}(\bar k_2).
    \end{align*}
    Thus, by combining \eqref{eq-bark1-k2-h-def} and \eqref{ieq-sigma1-2-3-design}
    with $\bar k_2> \frac{a}{b_a} $, we obtain \eqref{ieq-bark1-k2-h-sigma123}. 
    Consequently, Problem \ref{Problem-Robust_output_regulation_multiple_actuators} 
    is solved by the distributed dynamic output feedback control law \eqref{eq-distributed-output-feedback-control-law}.
\end{Proof}

\begin{Remark}\label{Remark-parameters-design-relationship}
	If $\mathrm{max}\{\phi_1(N\bar k_1),\frac{a}{b_a}\}=\phi_1(N\bar k_1)$, for instance, when $\frac{a}{b_a}\leq 0$, i.e., the actuators are strictly stable or marginally stable,
then \eqref{ieq-bark1-k2-h-sigma123} is simplified to
\begin{align*}
	&N\bar k_1>k^*_1, \quad \bar k_2>\phi_1(N\bar k_1)  \notag\\
	& \bar h>\mathrm{max}\{1, \phi_2(\bar k_2)\},  \quad   \mathrm{min}\{\sigma_1, \sigma_2\}>\sigma^{*}(\bar k_2).
\end{align*} 
As discussed in Remark \ref{Remark-design-bark1-k2-h}, 
a straightforward design for $\bar k_1$, $\bar k_2$, $\bar h$, $\sigma_1$, and $\sigma_2$,
can be expressed as
\begin{align*}
	 \bar k_1=\frac{k_1}{N}, \quad \bar k_2=k_2, \quad \bar h=h, \quad \mathrm{min}\{\sigma_1, \sigma_2\}>\sigma^{*}( \bar k_2)
\end{align*}
in which,  $k_1$, $k_2$, and $h$ satisfy \eqref{ieq-h-k1-k2-theorem1}.
On the other hand, if $\mathrm{max}\{\phi_1(N\bar k_1),\frac{a}{b_a}\}=\frac{a}{b_a}$, i.e., $\frac{a}{b_a}>0$, then one can design 
\begin{align*}
  &\bar k_1=\frac{k_1}{N}, \quad \bar k_2>\frac{a}{b_a} \notag\\
  &\bar h>\mathrm{max}\{1, \phi_2(\bar k_2)\}, \quad \mathrm{min}\{\sigma_1, \sigma_2\}>\sigma^{*}(\bar k_2).
\end{align*}
In other words, our approach can also handle exponentially unstable actuators.
\end{Remark}

\begin{Remark}
Cooperative control of parallel actuators for linear output regulation 
has been investigated in \cite{Lim_Oh-2024Auto} for a certain plant and 
in \cite{Xu_Su_Liu-2025Auto} for an uncertain plant. 
However, in these works, the design of feedback gains and observer parameters 
all depends on the number of actuators.
In contrast, the proposed distributed output feedback control law in \eqref{eq-distributed-output-feedback-control-law} significantly reduces this dependency. 
Specifically, both the distributed internal model \eqref{Internal-Model-i} and 
the high-gain observers in \eqref{eq-distributed-observer} 
are completely independent of the number of actuators, 
with only the single gain parameter $\bar k_1$ retaining this dependency. 
Moreover, while \cite{Lim_Oh-2024Auto} and \cite{Xu_Su_Liu-2025Auto} assume 
that the actuators are Hurwitz stable, our design eliminates this requirement.
\end{Remark}

\section{Numerical Examples}\label{Section-Example}
This section provides two numerical examples to illustrate the effectiveness of our design.

\subsection{Linear robust output regulation with strictly stable actuators }\label{subsection-simulation-E1}

\begin{figure}[H]
	\centering
	\includegraphics[width=0.8\linewidth, trim=0 0 0 30]{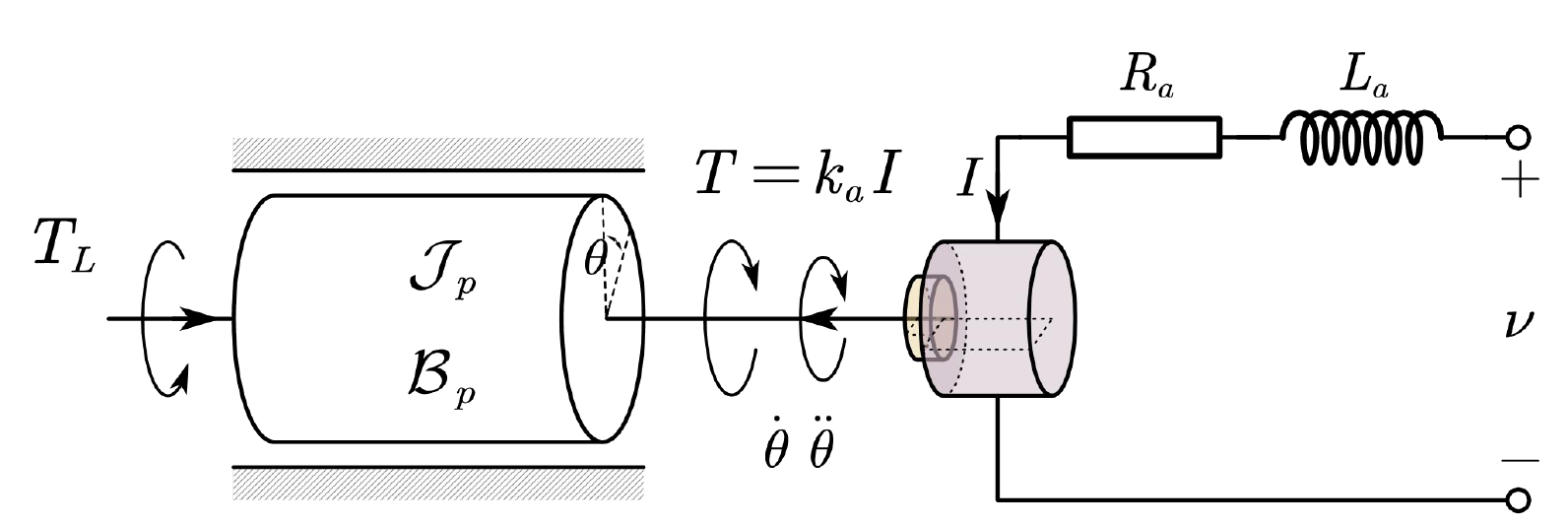}
	\caption{The motor-driven system of one electric motor driving an uncertain shaft.}\label{motor-driven system-1}
\end{figure}

We first consider the motor-driven system of one electric motor driving an uncertain shaft 
with a load torque as shown in Figure \ref{motor-driven system-1}. 
Denote the angle of the shaft, the load torque, and the electric torque provided 
by the electric motor by $\theta$, $T_L$, and $T$, respectively. Then, with the state $\mathrm{col}\left(\xi_1,\xi_2\right)=\mathrm{col}\left(\theta,\dot\theta\right)$, 
the input $u_p=T$, and the exogenous signal $v=\mathrm{col}\left(\theta_r, \dot\theta_r, T_L\right)$,
in which, $\theta_r$ is the reference angle, 
the motion of the uncertain shaft can be modeled as the following plant: 
\begin{align}\label{eq-simulation-plant-model}
	\dot \xi_1&=\xi_2  \notag\\
	\dot \xi_2&=\left(-\frac{\mathcal{B}_p}{\mathcal{J}_p}+w_1\right)\xi_2+\left(\frac{1}{\mathcal{J}_p}+w_2\right)u_p\notag\\
	&\quad+\begin{bmatrix} 0 & 0 & -\frac{1}{\mathcal{J}_p}+w_3\end{bmatrix}v \notag\\
	y&=\xi_1  \notag\\
	e&=y-\begin{bmatrix} 1 & 0 & 0 \end{bmatrix}v
\end{align}
where $\mathcal{J}_p$ and $\mathcal{B}_p$ denote the moment of inertia and the damping coefficient, respectively, 
and $w=\mathrm{col}\left(w_1,w_2,w_3\right)$ represents the uncertainties. 
Suppose $\mathcal{J}_p=0.5\,\mathrm{kg}{\cdot}\mathrm{m}^2$, $\mathcal{B}_p=1\,\mathrm{N}{\cdot}\mathrm{m}{\cdot}\mathrm{s/rad}$,
and $w\in \mathbb{W}=\{w: w_i\in[-0.3,0.3], i=1,2,3\}$.

Suppose the reference angle $\theta_{r}$ is a sinusoidal function with the frequency
$\omega=1\,\mathrm{rad/s}$
and the external load torque $T_{L}=2\,\mathrm{N}{\cdot}\mathrm{m}$. 
Then, the exogenous signal $v=\mathrm{col}\left(\theta_r, \dot\theta_r, T_L\right)$ 
can be generated by an exosystem of the form \eqref{eq-exosystem} with
\begin{align*}
	S=\begin{bmatrix} 0 & 1 &  0\\ -1 & 0 & 0\\ 0 & 0 & 0 \end{bmatrix}.
\end{align*}
Hence, Assumption \ref{Ass-nonnegative} is satisfied.

In addition, denote the resistance, the inductance, the torque constant, and the voltage input of 
the electric motor as $R_a$, $L_a$, $k_a$, and $\nu$, respectively. 
Then, with the state $x_1=T$ and input $u_1=\nu$, the actuator dynamics can be described by
\begin{align*}
	\dot x_1&=-\frac{R_{a}}{L_{a}}x_1+\frac{k_{a}}{L_{a}}u_1  \notag\\
	y_1&=x_1.
\end{align*}
Suppose $R_{a}=0.1\,\mathrm{\Omega}$, $L_{a}=0.01\,\mathrm{H}$, and $k_{a}=0.5\,\mathrm{N}{\cdot}\mathrm{m/A}$. 
It is clear that this actuator is strictly stable.

\begin{figure}[H]
	\centering
	\includegraphics[width=0.8\linewidth, trim=0 20 0 30]{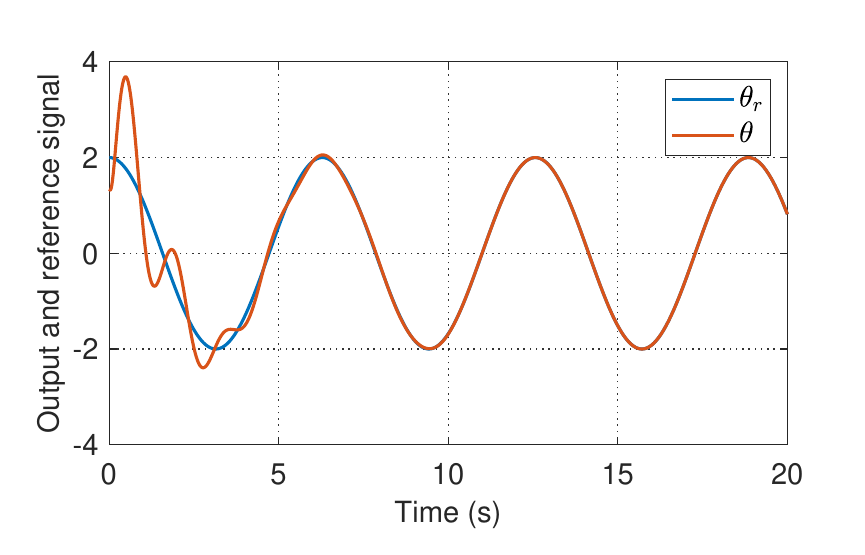}
		\caption{Plant output and reference signal under single motor operation.}\label{figE11-plant-output-single}
\end{figure}

By Theorem \ref{Thorem-single-actuator}, this robust output regulation problem with a single actuator 
can be solved by the dynamic output feedback control law of the form \eqref{eq-control-law-single-actuator}
and a specific design of all parameters in this control law is given below. 
Since the minimal polynomial of $S$ is $p(s)=s^3+s$, let 
\begin{align}\label{eq-simulation-Phi-Psi}
	\Phi=
	\begin{bmatrix}
		0  &  1  &  0 \\
		0  &  0  &  1 \\
		0  & -1  &  0
	\end{bmatrix},\quad
    \Psi=
    \begin{bmatrix}
    	1  \\  0  \\  0
    \end{bmatrix}^{T}.
\end{align} 
For convenience, we select the same controllable pairs
\begin{align}\label{eq-M_i_N_i_example}
	M_1=M_2=\begin{bmatrix}
		   0  &  1  &  0 \\
		   0  &  0  &  1 \\
		  -8  & -12 & -6
	   \end{bmatrix},\quad 
    N_1=N_2=\begin{bmatrix}
   	       0  \\  0  \\  1
        \end{bmatrix}.
\end{align}
By solving the Sylvester equations in \eqref{eq-Sylvester_equation} with 
$(\Phi,\Psi)$ in \eqref{eq-simulation-Phi-Psi} and $(M_i,N_i), i=1,2$, in \eqref{eq-M_i_N_i_example}, 
we have
\begin{align}\label{eq-simulation-T-E1}
	T_1=T_2=\begin{bmatrix}
		0.1250 & -0.0880  &  0.1090 \\
		0 &  0.0160  & -0.0880 \\
	   0 &  0.0880  &  0.0160
	\end{bmatrix} 
\end{align}
which yields 
\begin{align}\label{eq-simulation-PsiT-E1}
	\Psi T^{-1}_1=\Psi T^{-1}_2=\begin{bmatrix}
		0.1250 & -0.0880  &  0.1090
	\end{bmatrix}.
\end{align}
Further, select $\gamma_0=1$ and $\delta_0=\delta_1=4$ so that the polynomials $f(\lambda)$ and $g(\lambda)$, defined in \eqref{eq-stable_polynomial} and \eqref{eq-def-g(s)}, respectively, are stable.
It follows that
\begin{align}\label{eq-simulation-A0h-E1}
	A_0( h)=\begin{bmatrix}
		-4 h  &  1  \\  -4 h^2  &  0 
	\end{bmatrix}, \quad
    B_0(h)=\begin{bmatrix}
    	4 h  \\  4 h^2
    \end{bmatrix}.
\end{align}
Then, we can obtain a dynamic output feedback control law \eqref{eq-control-law-single-actuator} 
by choosing $ k_1=2$, $ k_2=6$, and $ h=14$ that satisfy \eqref{ieq-h-k1-k2-theorem1}. 
The simulation is conducted with random initial conditions from the range $[-3,3]$. 
Figure \ref{figE11-plant-output-single} demonstrates the tracking performance of the plant output $\theta$ to the reference signal $\theta_r$.

\begin{figure}[H]
	\centering
	\includegraphics[width=0.8\linewidth]{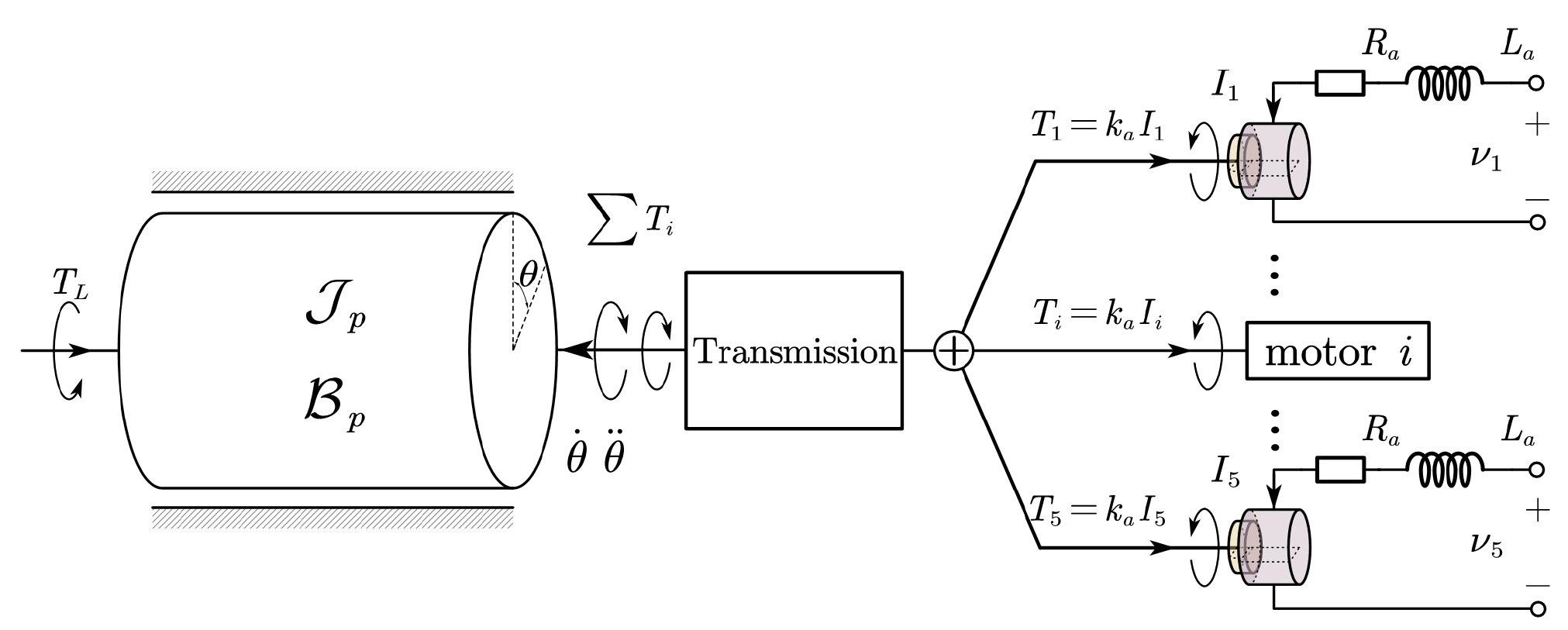}
	\caption{The motor-driven system of five electric motors collectively driving a common uncertain shaft.}\label{motor-driven system-2}
\end{figure}

Next, we consider the case where the uncertain shaft  is collectively driven by five electric motors as shown in Figure \ref{motor-driven system-2}. 
In this case, the plant is modeled as in \eqref{eq-simulation-plant-model} 
by only replacing $u_p=T$ with $u_p=\sum_{i=1}^{5}T_i, i=1,\dots, 5$, 
where $T_i$ is the electric torque of the $i$th electric motor. 
Assume that the five electric motors have the same resistance $R_a$, inductance $L_a$, 
and the torque constant $k_a$. 
Then, with the state $x_i=T_i$ and the input $u_i=\nu_i$,
where $\nu_i$ is the voltage input of the $i$th electric motor, 
these multiple actuators can be described by
\begin{align*}
	\dot x_{i}&=-\frac{R_{a}}{L_{a}}x_{i}+\frac{k_{a}}{L_{a}}u_{i}  \notag\\
	y_{i}&=x_{i}, \quad i=1,\ldots, 5. 
\end{align*}

\begin{figure}[H]
	\centering
	\includegraphics[width=0.7\linewidth,  trim=0 0 0 10]{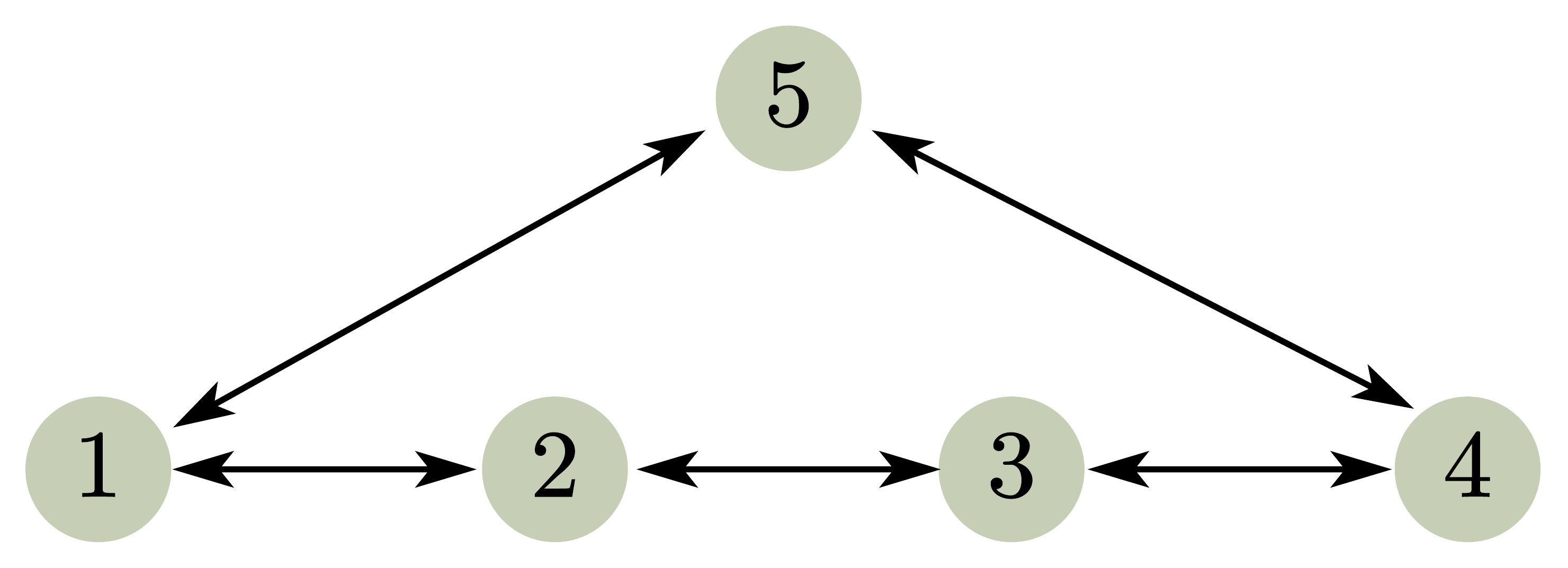}
	\caption{Communication network of five actuators.}\label{Graph-U}
\end{figure}

Assume the same exosystem and suppose $\mathcal{J}_p$, $\mathcal{B}_p$, $R_{a}$, $L_{a}$, and $k_{a}$ 
are the same as in the single electric motor case. 
The communication network of the five electric motors is depicted in Figure \ref{Graph-U}.
Then, Assumptions \ref{Ass-nonnegative}  and   \ref{Ass-connected-graph} are satisfied.

\begin{figure}[H]
	\centering
	\includegraphics[width=0.8\linewidth,  trim=0 20 0 30]{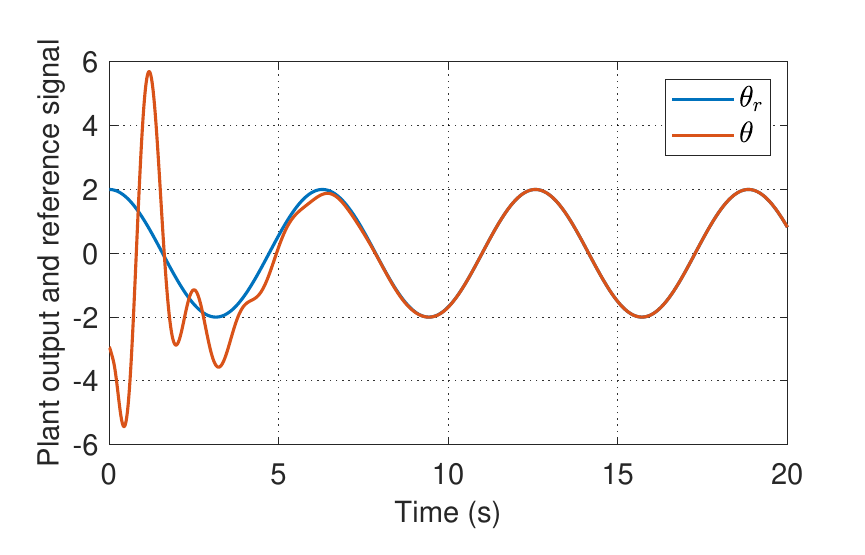}
	\caption{Plant output and reference signal with cooperative parallel operation of five electric motors.}\label{figE12-plant-output-multiple}
\end{figure}

\begin{figure}[H]
	\centering
	\includegraphics[width=0.8\linewidth,  trim=0 20 0 30]{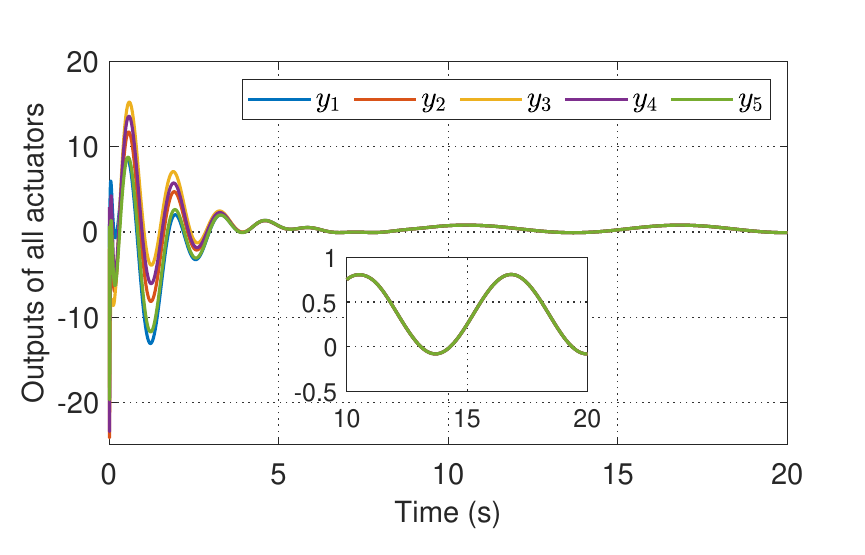}		
\caption{Plant input provided by five electric motors.}\label{figE13-input-sharing}
\end{figure}

By Remark \ref{Remark-parameters-design-relationship}, 
since $\frac{a}{b_a}=-0.2<0$, one can simply design 
$
	\bar k_1=\frac{k_1}{5}=0.4, \bar k_2=k_2=6, \bar h=h=14
$,
and the other parameters as in \eqref{eq-simulation-Phi-Psi} to \eqref{eq-simulation-A0h-E1}.
Then, we can obtain a distributed dynamic output feedback control law   \eqref{eq-distributed-output-feedback-control-law} 
by further choosing $\sigma_1=\sigma_2=1$. 
The simulation is performed with random initial conditions from the range $[-3,3]$. 
Figure \ref{figE12-plant-output-multiple} demonstrates the tracking performance of the plant collectively driven by  the five electric motors and Figure \ref{figE13-input-sharing} shows that the plant input~$u_p$ 
is equally shared by the outputs $y_i, i=1,\ldots, 5$ of the five electric motors.

\subsection{Linear robust output regulation with unstable actuators}

We next consider the following uncertain linear minimum-phase plant from \cite{Liu-Huang-2017IJRNC}:
\begin{align}\label{eq-simulation-plant-ex2}
	\dot z&=(-6+w_1)z+(3+w_2)\xi_1+\begin{bmatrix}	1+w_7 & 0	\end{bmatrix}
	\begin{bmatrix}
		v_1 \\ v_2
	\end{bmatrix} \notag\\
    \dot\xi_1&=\xi_2 \notag\\
    \dot\xi_2&=(4+w_3)z+(-20+w_4)\xi_1+(-9+w_5)\xi_2  \notag \\
    &\quad +\begin{bmatrix} 0 & 1+w_8 \end{bmatrix}
    \begin{bmatrix} v_1 \\ v_2 \end{bmatrix}+(2+w_6)u_p \notag\\
    y&=\xi_1  
\end{align}
where $w=\mathrm{col}\left(w_1,\dots,w_8\right)\in\mathbb{W}$ 
with $\mathbb{W}=\{w: w_i\in[-0.5,0.5],i=1,\ldots,8\}$, and the following exosystem:
\begin{align*}
	\begin{bmatrix}	\dot v_1 \\ \dot v_2\end{bmatrix}
	=\begin{bmatrix} 0 & 1 \\ -1 & 0 \end{bmatrix}
	\begin{bmatrix}	 v_1 \\  v_2\end{bmatrix}, \quad
	y_0=v_1.
\end{align*}
Clearly, Assumption \ref{Ass-nonnegative} is satisfied.

Different from \cite{Liu-Huang-2017IJRNC} where the plant input $u_p$ 
is regarded as the control input, 
we first consider the case where $u_p$ is provided by the following single actuator:
\begin{align*}
	\dot x_1&=x_1+10u_1 \notag\\
	       y_1&=x_1      \notag\\
	     u_p&=y_1.
\end{align*}
It is easy to see that this actuator is unstable. Such a model can be used to characterize actuators in active suspension systems operating under degraded conditions.

\begin{figure}
	\centering
	\includegraphics[width=0.8\linewidth,  trim=0 20 0 30]{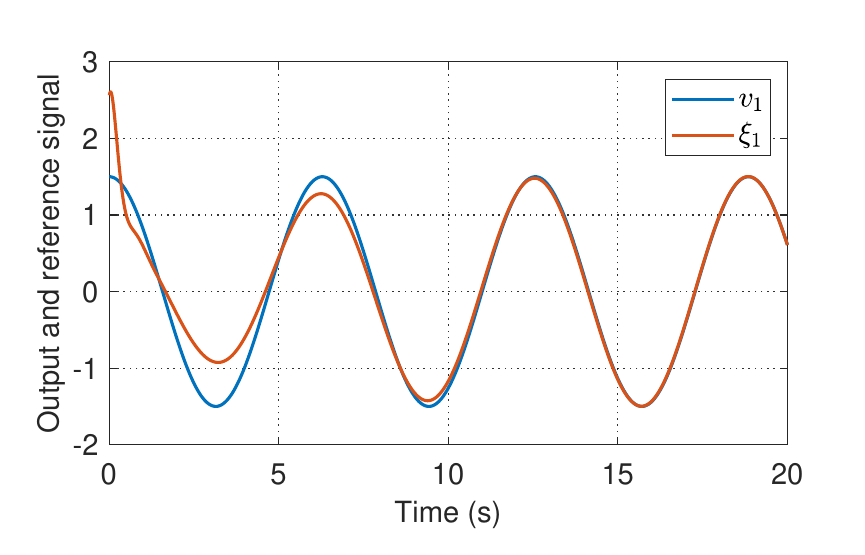}
	\caption{Plant output and reference signal under single actuator operation}\label{figE21-plant-output-single}
\end{figure}

Following the same design process from \eqref{eq-simulation-Phi-Psi} to \eqref{eq-simulation-PsiT-E1} 
in Subsection \ref{subsection-simulation-E1}, 
we can design
\begin{align}\label{eq-simulation-parameters-ex2}
	\Phi&=\begin{bmatrix} 0  &  1 \\ -1 &  0 \end{bmatrix},\quad
	\Psi=\begin{bmatrix} 1  \\  0  \end{bmatrix}^{T} \notag \\
	M_1&=M_2=\begin{bmatrix} 0  &  1  \\ -1  &  -2 \\ \end{bmatrix},\quad 
	N_1=N_2=\begin{bmatrix} 0  \\  1 \end{bmatrix}  \notag\\
	T_1&=T_2=\begin{bmatrix} 0 & -0.5 \\ 0.5 &  0 \end{bmatrix}, \quad 
	\Psi T^{-1}_1=\Psi T^{-1}_2=\begin{bmatrix} 0  &  2  \end{bmatrix}. 
\end{align}
By selecting  $\gamma_0=1$ and $\delta_0=\delta_1=4$,
we have the same $A_0(h)$ and $B_0(h)$ as in \eqref{eq-simulation-A0h-E1}. 
Then, we can obtain a dynamic output feedback control law \eqref{eq-control-law-single-actuator}
by choosing $k_1=2$, $k_2=3$, and $h=5$ that satisfy \eqref{ieq-h-k1-k2-theorem1}. 
The simulation is performed with random initial conditions from the range $[-3,3]$. 
It can be observed from Figure \ref{figE21-plant-output-single}
that plant output asymptotically tracks the reference signal.

\begin{figure}[H]
	\centering
	\includegraphics[width=0.8\linewidth,  trim=0 20 0 30]{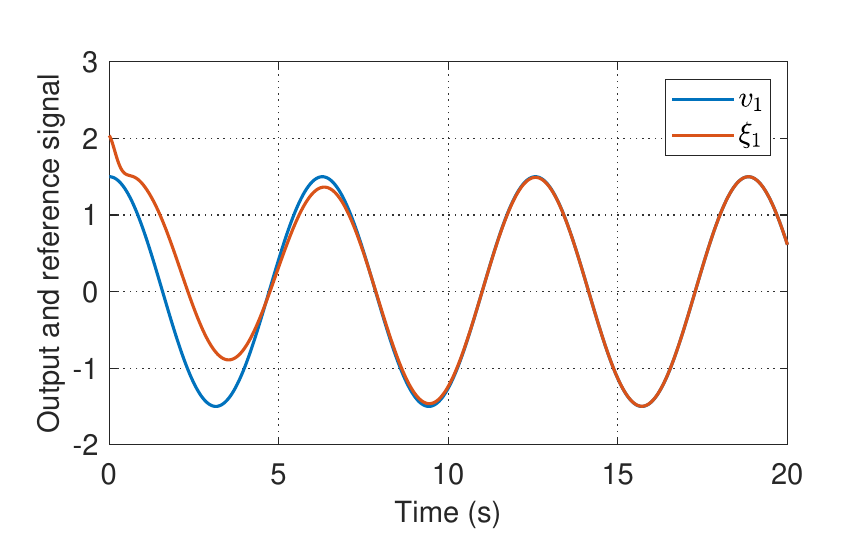}
	\caption{Plant output and reference signal with cooperative parallel operation of five actuators.}\label{figE22-plant-output-multiple}
\end{figure}

\begin{figure}[H]
	\centering
	\includegraphics[width=0.8\linewidth,  trim=0 20 0 30]{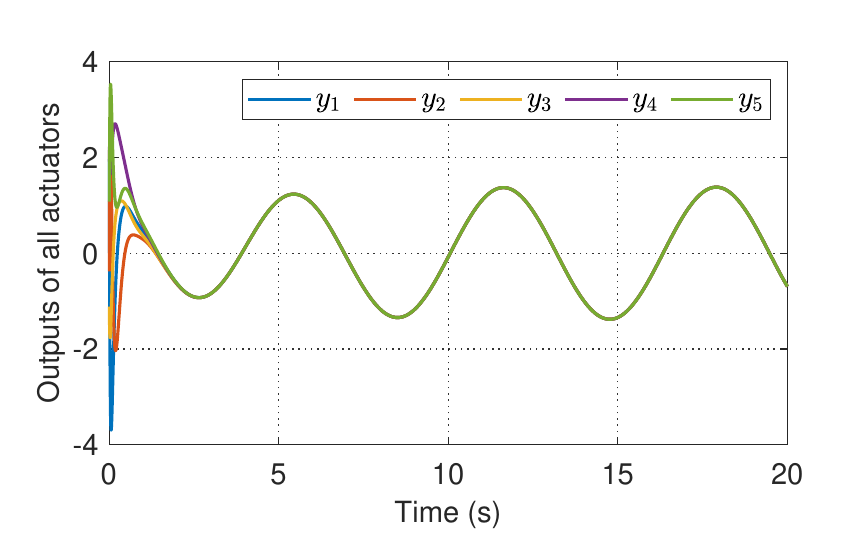}
	\caption{Plant input provided by five actuators.}\label{figE23-input-sharing}
\end{figure}

We next consider the case where the uncertain plant \eqref{eq-simulation-plant-ex2} 
is driven by five actuators as follows:
\begin{align*}
	\dot x_i&=x_i+10u_i \\
	y_i&=x_i, \quad i=1,\dots, 5    \\
	u_p&=\sum_{i=1}^{5}y_i.
\end{align*}
With some computations, we can design
$
	\bar k_1=\frac{k_1}{5}=0.4, \bar k_2=3.5, \bar h=5.5
$.
Further, from Lemma \ref{lemma-sigma-design}, 
by selecting $\sigma_1=2$ and $\sigma_2=3$, 
we can obtain the distributed dynamic output feedback control law \eqref{eq-distributed-output-feedback-control-law} with the remaining parameters designed as in \eqref{eq-simulation-parameters-ex2}. 
The simulation is performed with random initial condition from the range $[-3, 3]$ 
and the communication network  in Figure \ref{Graph-U}. 
Figure \ref{figE22-plant-output-multiple} shows the evolution of the plant output and the reference signal with cooperative parallel operation of the five actuators. 
Figure \ref{figE23-input-sharing} further shows that the plant input $u_p$ is equally shared by the outputs $y_i, i=1,\dots, 5$ of the five actuators.

\begin{figure}[H]
	\centering
	\includegraphics[width=0.8\linewidth,  trim=0 20 0 30]{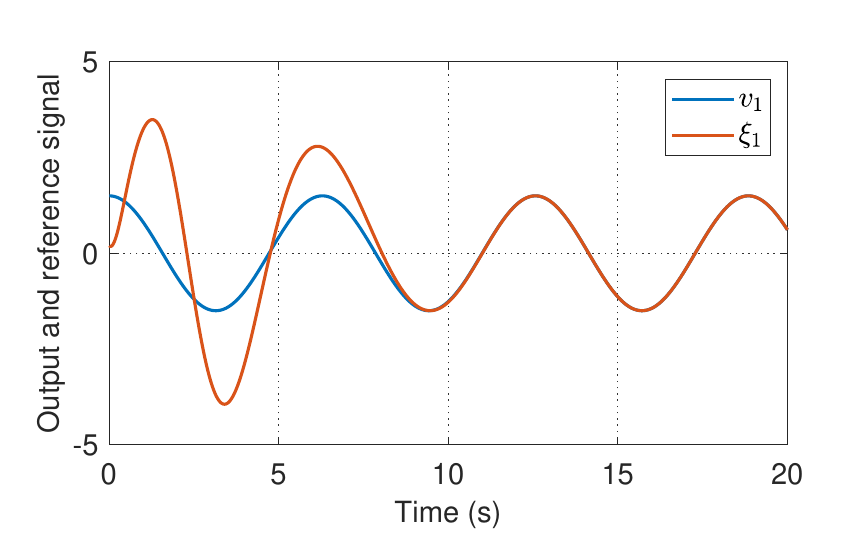}
		\caption{Plant output and reference signal using the control law of \cite{Lim_Oh-2024Auto}.}\label{figE31-plant-output-multiple}
\end{figure}

\begin{figure}[H]
	\centering
		\includegraphics[width=0.8\linewidth,  trim=0 20 0 30]{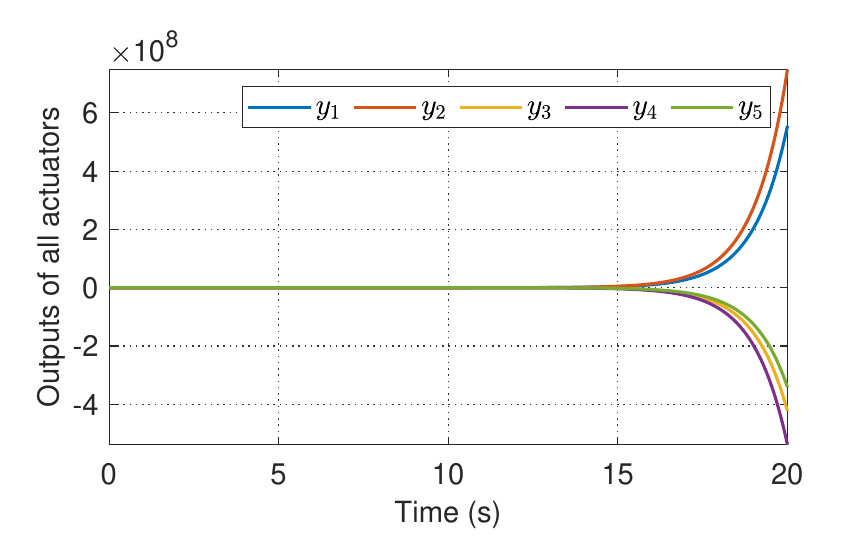}
		\caption{Plant input provided by five actuators using the control law of \cite{Lim_Oh-2024Auto}.}\label{figE32-input-sharing}
\end{figure}

\begin{figure}[H]
	\centering
\includegraphics[width=0.8\linewidth, trim=0 20 0 30]{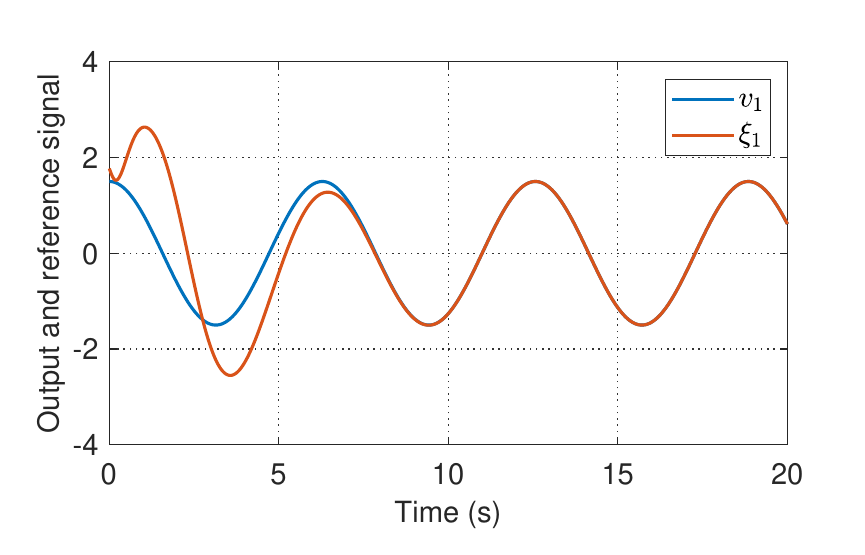}
		\caption{Plant output and reference signal using the control law of \cite{Xu_Su_Liu-2025Auto}.}\label{figE33-plant-output-multiple}
\end{figure}

\begin{figure}[H]
	\centering
	\includegraphics[width=0.8\linewidth,  trim=0 20 0 30]{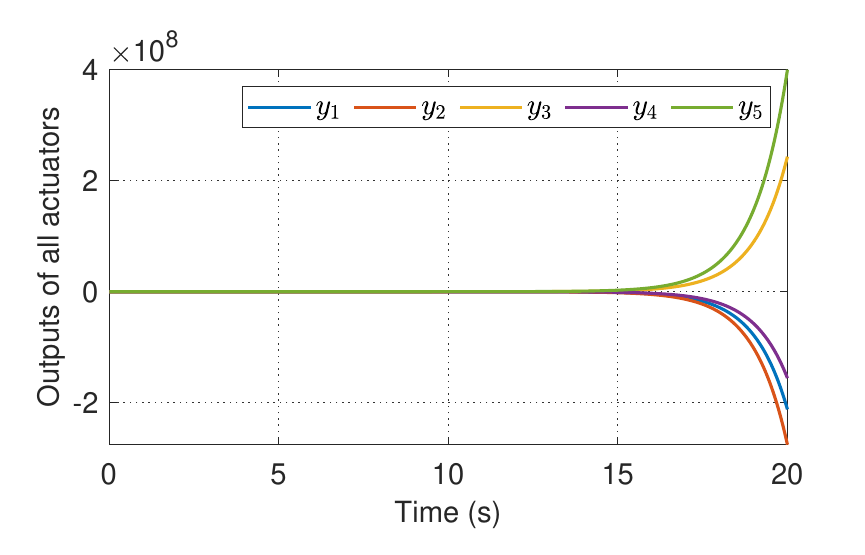}
		\caption{Plant input provided by five actuators using the control law of \cite{Xu_Su_Liu-2025Auto}.}\label{figE34--input-sharing}
\end{figure}

For comparison, we also conduct simulations 
using the control laws from \cite{Lim_Oh-2024Auto} and \cite{Xu_Su_Liu-2025Auto} with random initial conditions from the range $[-3,3]$. 
For simplicity, we let $w_i=0, i=1,\dots, 8$. 
The simulation results are shown in Figures \ref{figE31-plant-output-multiple} to \ref{figE34--input-sharing}. 
Specifically, Figures \ref{figE31-plant-output-multiple} and \ref{figE33-plant-output-multiple} demonstrate that output regulation can be achieved by the distributed control laws from \cite{Lim_Oh-2024Auto} and \cite{Xu_Su_Liu-2025Auto}. 
However, as indicated in Figures \ref{figE32-input-sharing} and \ref{figE34--input-sharing}, 
input sharing among the five unstable actuators is not maintained using these control laws.

\section{Conclusion}\label{Section-Conclusions}
In this paper, we have investigated cooperative control of parallel actuators 
for robust output regulation of an uncertain linear minimum-phase plant. 
We have first examined the special case where the uncertain linear plant is driven by a single linear actuator. 
Leveraging the internal model approach, 
we have solved this special problem 
by designing a dynamic output feedback control law. 
We have then extended this result to cooperative control of parallel actuators, developing a distributed dynamic output feedback control law that is nearly independent of the number of actuators. 
Furthermore, we have removed the restrictive Hurwitz stability assumption 
on the actuators and have demonstrated that 
robust output regulation of the closed-loop system and plant input sharing 
among the actuators can be achieved over undirected communication networks.
Future research will explore more general and practical scenarios involving uncertain actuators and multi-input multi-output uncertain plants \cite{Zhou-2025TAC}. In addition, addressing transient performance constraints on actuators and developing adaptive mechanisms to dynamically adjust the gains constitute compelling directions for future investigation.

\appendices
\section{A Technical Lemma}\label{Appendix-lemma}

\begin{Lemma}\label{Lemma_asymptotically_stable_system}
Consider the following uncertain linear system:
\begin{align}\label{eq-matrix_lemma}
	\begin{bmatrix}
		\dot x  \\ \dot y
	\end{bmatrix}
	=\begin{bmatrix}
		M_1(w,\mu_1) & N_1(w, \mu_1) \\
		N_2(w, \mu_1) & N_3(w, \mu_1)+\mu_2M_2(w)
	\end{bmatrix}\begin{bmatrix}
		x  \\  y
	\end{bmatrix}
\end{align}
where $x\in \RR^{n}$, $y\in \RR^{m}$,
$\mu_1\in\RR$, $w \in \WW$ with $\WW$ being a compact set, 
$M_2(w) \in \RR^{m \times m}$ is Hurwitz, and 
\begin{align}\label{ieq-bounded_Ni_lemma}
	\Vert N_i(w, \mu_1)\Vert \leq \varphi_i(\mu_1), \quad \forall\, w \in \WW, \quad i=1,2,3
\end{align}
with $\varphi_i(\cdot)$ being some positive functions. 
Suppose there exists $\mu^{*}_1>0$
such that $M_1(w,\mu_1) \in \RR^{n \times n}$ is Hurwitz for any $\mu_1>\mu^{*}_1$.
Then, there exists a positive function $\phi(\mu_1)$ defined for $\mu_1>\mu^{*}_1$, 
such that system \eqref{eq-matrix_lemma} is asymptotically stable 
for all $w \in \WW$ when $\mu_1>\mu^{*}_1$ and either of the following two conditions holds:
\begin{enumerate}[(H1)]
	\item $\mu_2\in\RR$ satisfies $\mu_2>\phi(\mu_1)$; \label{eq-lemma-condition-1}
	\item $M_2(w)=\alpha(w) I_m$ with $\alpha(w)<0$ and $\mu_2=\mathrm{diag}\left\{\mu_{21}, \dots, \mu_{2m}\right\}\in\RR^{m\times m}$ satisfies 
$\min\left\{\mu_{2i} : i=1,\ldots, m \right\}>\phi(\mu_1)$. \label{eq-lemma-condition-2}
	\end{enumerate}
\end{Lemma}

\begin{Proof}
	First, we prove that system \eqref{eq-matrix_lemma} is asymptotically stable 
for all $w \in \WW$ when
$\mu_1>\mu^{*}_1$ and $\mu_2$ satisfies condition (H\ref{eq-lemma-condition-1}). 

	Since $M_1(w,\mu_1)$ is Hurwitz for any $\mu_1>\mu^{*}_1$, 
there exists a positive definite matrix $P(w,\mu_1)$ such that 
	\begin{align}\label{ieq-like_Lyapunov_equation}
		M^{T}_1(w,\mu_1)P(w,\mu_1)+P(w,\mu_1)M_1(w,\mu_1)= -I_n. 
	\end{align}
Similarly, since $M_{2}(w)$ is Hurwitz, 
there also exists a positive definite matrix $Q(w)$ such that 
\begin{align}\label{ieq-Lyapunov_equation}
	M^{T}_2(w)Q(w)+Q(w)M_2(w)= -I_m.
\end{align}

Consider the following Lyapunov function candidate:
\begin{align*}
	V(x,y)=x^{T}P(w,\mu_1)x+y^{T}Q(w)y
\end{align*}
with $\mu_1>\mu^{*}_1$. 
Let 
\begin{align*}
	\underline{\alpha}(\mu_1)&=\min\limits_{w\in\WW}\{\lambda_{\min}(P(w,\mu_1)), \lambda_{\min}(Q(w))\}>0  \\
	\overline{\alpha}(\mu_1)&=\max\limits_{w\in\WW}\{\lambda_{\max}(P(w,\mu_1)), \lambda_{\max}(Q(w))\}>0.
\end{align*}
Since  $\underline{\alpha}(\mu_1)\Vert \mathrm{col}(x,y) \Vert^2 \leq V(x,y)\leq \overline{\alpha}(\mu_1)\Vert \mathrm{col}(x,y) \Vert^2$,  
$V(x,y)$ is a positive definite function for all $w\in\WW$.  
Using \eqref{ieq-like_Lyapunov_equation} and \eqref{ieq-Lyapunov_equation}, 
the time derivative of $V(x,y)$ along the trajectory of system \eqref{eq-matrix_lemma} satisfies
\begin{align}\label{eq-dot_V_xy}
	&\quad\ \dot V(x,y)\notag \\
&= -x^{T}x -\mu_2y^{T}y  + 2y^{T}\left(N^{T}_3(w,\mu_1)Q(w)\right)y  \notag\\
	&\quad +2x^{T}\left(P(w,\mu_1)N_{1}(w,\mu_1)+N^{T}_2(w,\mu_1)Q(w)\right)y.
\end{align}
In particular, using \eqref{ieq-bounded_Ni_lemma} and Young's inequality, 
we have
\begin{align}\label{ieq-yQNy}
	 2y^{T}\left(N^{T}_3(w,\mu_1)Q(w)\right)y \leq  2\varphi_3(\mu_1)\Vert Q(w) \Vert y^{T}y
\end{align}
\begin{align}\label{ieq-xPNy}
	&\quad\ 2x^{T}P(w,\mu_1)N_{1}(w,\mu_1)y  \notag\\
	&\leq  2\varphi_1(\mu_1)\Vert x \Vert \Vert P(w,\mu_1) \Vert \Vert y \Vert \notag\\
	&\leq  \frac{1}{4}x^{T}x + 4\varphi^2_1(\mu_1)\Vert P(w,\mu_1) \Vert^2y^{T}y
\end{align}
and
\begin{align}\label{ieq-xNQy}
	&\quad\ 2x^{T}N^{T}_2(w,\mu_1)Q(w)y  \notag\\
	&\leq  2\varphi_2(\mu_1)\Vert x \Vert \Vert Q(w) \Vert \Vert y \Vert \notag\\
	&\leq  \frac{1}{4}x^{T}x + 4\varphi^2_2(\mu_1)\Vert Q(w) \Vert^2y^{T}y.  
\end{align}
Putting together \eqref{eq-dot_V_xy} to \eqref{ieq-xNQy}, we obtain
\begin{align}\label{eq-derivate-V(xy)}
&	\quad\ \dot V(x,y)\notag\\
&\leq-\frac{1}{2}x^{T}x -\mu_2y^{T}y +2\varphi_ 3(\mu_1)\Vert Q(w) \Vert y^{T}y  \notag\\
	&\quad  +4\left(\varphi^2_1(\mu_1)\Vert P(w,\mu_1) \Vert^2 +\varphi^2_2(\mu_1)\Vert Q(w) \Vert^2\right)y^{T}y   \notag\\
	&= -\frac{1}{2}x^{T}x-\big(\mu_2-2\varphi_3(\mu_1)\Vert Q(w) \Vert \notag\\
& \quad  -4\left(\varphi^2_1(\mu_1)\Vert P(w,\mu_1) \Vert^2 +\varphi^2_2(\mu_1)\Vert Q(w) \Vert^2\right)\big)y^{T}y.
\end{align}
Define
\begin{align}\label{eq-phi-function-def}
	\phi(\mu_1)&:=\max\limits_{w\in\WW}\big\{2\varphi_3(\mu_1)\Vert Q(w)\Vert+ 4\left(\varphi^2_1(\mu_1)\Vert P(w,\mu_1) \Vert^2 \right. \notag\\
&\quad\ \left. +\varphi^2_2(\mu_1)\Vert Q(w) \Vert^2\right)\big\}.
\end{align}
Then, for any $\mu_2>\phi(\mu_1)$ with $\mu_1>\mu^{*}_1$, we have
\begin{align}\label{eq-derivative-V(xy)-C1}
	\dot V(x,y)
	\leq -\beta(x^{T}x+y^{T}y)
\end{align}
where $\beta =\min\left\{\frac{1}{2}, \mu_2-\phi(\mu_1)\right\}>0$. 
Therefore, system \eqref{eq-matrix_lemma} is asymptotically stable for all $w\in\WW$.

Next, we further show that system \eqref{eq-matrix_lemma} is also asymptotically stable 
for all $w \in \WW$ when $M_2(w)$ and $\mu_2$ satisfy condition (H\ref{eq-lemma-condition-2}). In this case, we let
\begin{align}\label{eq-Q-def}
	Q(w)=-\frac{1}{2\alpha(w)}I_m
\end{align}
which satisfies \eqref{ieq-Lyapunov_equation}. 
Then, using the same arguments as those in the case of condition (H\ref{eq-lemma-condition-1}), 
\eqref{eq-dot_V_xy} is updated by replacing
$-\mu_2y^{T}y$ with 
$y^{T}\left(M^{T}_2(w)\mu_2Q(w)+Q(w)\mu_2M_2(w)\right)y$ where $Q(w)$ is defined in \eqref{eq-Q-def}. 
Recall the definitions $M_2(w)$ and $\mu_2$ in condition (H\ref{eq-lemma-condition-2}). 
We have
\begin{align*} 
	y^{T}\left(M^{T}_2(w)\mu_2Q(w)+Q(w)\mu_2M_2(w)\right)y = -y^{T}\mu_2y.
\end{align*}
Hence, further using \eqref{ieq-yQNy} to \eqref{ieq-xNQy}, 
the time derivative of $V(x,y)$ in \eqref{eq-derivate-V(xy)} is updated to
\begin{align*} 
  	&\quad\ \dot V(x,y) \\
   	&\leq -\frac{1}{2}x^{T}x-y^{T}\Big(\mu_2- 2\varphi_3(\mu_1)\Vert Q(w) \Vert I_m \notag\\
   	&\quad -4\left(\varphi^2_1(\mu_1)\Vert P(w,\mu_1) \Vert^2+\varphi^2_2(\mu_1)\Vert Q(w) \Vert^2\right)I_m\Big)y.
\end{align*}
Keep the definition of $\phi(\mu_1)$ in \eqref{eq-phi-function-def} with $Q(w)$ 
given by \eqref{eq-Q-def}.
Then, if $\min\{\mu_{2i} : i=1,\ldots, m \}>\phi(\mu_1)$ with $\mu_1>\mu^{*}_1$, 
the inequality  \eqref{eq-derivative-V(xy)-C1} holds with $\beta =\min\left\{\frac{1}{2}, \min\{\mu_{2i} : i=1,\ldots, m\}-\phi(\mu_1)\right\}>0$.
Therefore, system \eqref{eq-matrix_lemma} is asymptotically stable for all $w\in\WW$.	
\end{Proof}

\section{Proof of Lemma \ref{Result_of_state_feedback} }
\begin{Proof}
The closed-loop system composed of the augmented system \eqref{eq-augmented_system_transformed} and the static state feedback control law \eqref{eq-control_law_augmented_system} is 
\begin{align}\label{eq-closed_loop_augmented_system}
	\dot{\bar z} & = A_1(w) \bar z+A_2(w)D \bar\xi   \notag\\
	\dot{\bar\xi}& =\Lambda\bar\xi+G \zeta_1   \notag\\
	\dot{\zeta}_1 & =A_3(w) \bar{z}+C(w)\bar\xi+\tilde c(w,k_1)\zeta_1+b(w)\Psi T^{-1}_1\tilde\eta_1 \notag\\
	&\quad +b(w)\zeta_2    \notag\\
	\dot{\tilde \eta}_1 & =M_1\tilde\eta_1+\check{A}_3(w)\bar z+\check {C}_1(w)\bar\xi+\check{C}_2(w)\zeta_1 \notag\\
	\dot\zeta_2 & =\bar A_3(w,k_1) \bar{z}+\bar C_1(w,k_1)\bar\xi+\bar c_1(w,k_1)\zeta_1+\bar C_2(w,k_1)\tilde\eta_1 \notag\\
	&\quad +\bar c_2(w,k_1)\zeta_2+b_a\Psi T^{-1}_2\tilde\eta_2-b_ak_2\zeta_2 \notag\\
	\dot{\tilde \eta}_2 & =M_2 \tilde \eta_2+\hat A_3(w,k_1)\bar z+\hat C_1(w,k_1)\bar\xi+\hat C_2(w,k_1)\zeta_1 \notag\\
	&\quad +\hat C_3(w,k_1)\tilde\eta_1+\hat C_4(w,k_1)\zeta_2 .
\end{align}
Let $X=\mathrm{col}\left(\bar\xi, \bar z, \tilde\eta_1, \zeta_1, \tilde\eta_2, \zeta_2\right)$. 
Then, the closed-loop system~\eqref{eq-closed_loop_augmented_system} 
can be put into the following compact form:
\begin{align}\label{eq-closed_loop_system_compacted_form}
	\dot X=\bar A(w,k_1,k_2)X
\end{align}
where
\begin{align}\label{eq-bar-A-w-k-1-2-def}
	\bar A(w,k_1,k_2)=
	\left[\begin{array}{c c}
		\bar A_1(w,k_1) & \bar B_1(w,k_1) \\
		\bar B_2(w,k_1) & \bar B_3(w,k_1)+k_2\bar A_2(w)   
	\end{array}\right]
\end{align}
with the matrices 	
\begin{align}\label{eq-parameters-bar-Bi}
	&\bar A_1(w,k_1)=\notag\\
	&\begin{bmatrix}
		\Lambda & \zero & \zero & G & \zero  \\ 
		A_2(w)D & A_1(w) & \zero & \zero & \zero  \\
		\check{C}_1(w) & \check{A}_3(w) & M_1 & \check{C}_2(w) & \zero \\
		C(w) & A_3(w) & b(w)\Psi T^{-1}_1 & \tilde c(w,k_1) & \zero \\
		\hat C_1(w,k_1) & {\hat A}_3(w,k_1) & \hat C_3(w,k_1) & \hat C_2(w,k_1) & M_2 
	\end{bmatrix} \notag\\
 &\bar B_1(w,k_1)=\begin{bmatrix} \zero & \zero & \zero & b(w) & \hat C^{T}_4(w,k_1)  \end{bmatrix}^{T} \notag\\
 &\bar B_2(w,k_1)=\notag\\
 &\begin{bmatrix}	
 	{\bar C}_1(w,k_1) & {\bar A}_3(w,k_1) & {\bar C}_2(w,k_1) & {\bar c}_1(w,k_1) & b_a\Psi T^{-1}_2
 \end{bmatrix} \notag\\
&\bar B_3(w,k_1) ={\bar c}_2(w,k_1)\notag\\
&~~~~\bar A_2(w)=-b_a.
\end{align}

Next, we first show that there exists $k^{*}_1>0$ such that, 
for any $k_1>k^{*}_1$ and any $w\in\WW$, 
the matrix $\bar A_1(w,k_1)$ is Hurwitz. 
For this purpose, we rewrite $\bar A_1(w,k_1)$ as follows:
	\begin{align}\label{eq-transformed_nominal__system_matrix}
		\bar A_1(w,k_1)=\left[\begin{array}{cc|c}
			\bar A_{11}(w) & \bar A_{12}(w) & \zero \\
			\bar A_{21}(w) & \bar A_{22}(w)-k_1 b(w) & \zero \\
			\hline
			\bar A_{31}(w,k_1) & \hat C_2(w,k_1) & M_2 
		\end{array}\right]
	\end{align}
	where 
	\begin{align*}
		\bar A_{11}(w)&=
		\begin{bmatrix}
			\Lambda & \zero & \zero  \\ 
			A_2(w)D & A_1(w) & \zero \\
			\check{C}_1(w) & \check{A}_3(w) & M_1
		\end{bmatrix}, 	\
		\bar A_{12} =\begin{bmatrix}
			G \\ \zero \\ \check{C}_2(w)
		\end{bmatrix} \\ 
	    \bar A_{21}(w) &=\begin{bmatrix}
			C(w) & A_3(w) & b(w)\Psi T^{-1}_1
		\end{bmatrix} \\
		\bar A_{22}(w) & =c_r(w)+\gamma_{r-2}+\Psi T^{-1}_1N_1 \\
		 \bar A_{31}(w) &=\begin{bmatrix}
			\hat C_1(w,k_1) & {\hat A}_3(w,k_1) & \hat C_3(w,k_1)
		\end{bmatrix}.
	\end{align*}
From the definition of $\Lambda$ in \eqref{eq-augmented_system_transformed}, 
where the parameters $\gamma_{s}, s=0,\dots, r-2$,
are selected to ensure stability of the polynomial $f(\lambda)$ in \eqref{eq-stable_polynomial}, we know that
$\Lambda$ is Hurwitz. 
Since the plant \eqref{eq-plant} is assumed minimum-phase,
$A_1(w)$ is Hurwitz for all $w\in\WW$.
Moreover, by our design of the internal model,
the matrix $M_1$ is Hurwitz. 
Thus, the matrix $\bar A_{11}(w)$ is Hurwitz for all $w\in\WW$. 
In addition, note that $\Vert \bar A_{12}(w) \Vert $, $\Vert \bar A_{21}(w) \Vert$, and $\Vert \bar A_{22}(w)\Vert$ are bounded, and that $b(w)>0$. Then, by Lemma \ref{Lemma_asymptotically_stable_system}, there exists $k^{*}_1>0$ such that the following matrix: 
\begin{align*}
	\begin{bmatrix}
		\bar A_{11}(w)  &  \bar A_{12}(w) \\ 
		\bar A_{21}(w)  &  \bar A_{22}(w)-k_1 b(w)
	\end{bmatrix}
\end{align*}
is Hurwitz for any $k_1>k^{*}_1$ and for all $w\in\WW$. 
Further, since the matrix $M_2$ is also Hurwitz by design,
the matrix $\bar A_1(w,k_1)$ in \eqref{eq-transformed_nominal__system_matrix} 
is Hurwitz for any $k_1>k^{*}_1$ and for all $w\in\WW$.

Now, from \eqref{eq-parameters-bar-Bi}, we see that each entry of $\bar B_1(w,k_1)$, $\bar B_2(w,k_1)$, and $\bar B_3(w,k_1)$ is a polynomial function of $k_1$. 
Hence, $\Vert \bar B_i(w,k_1) \Vert , i=1,2,3$, 
can be bounded by some positive functions of $k_1$. 
Moreover, $\bar A_2(w)=-b_a <0$. 
Thus, invoking Lemma \ref{Lemma_asymptotically_stable_system} once more, 
there exists a positive function $\phi_1(k_1)$ such that 
for any $k_2>\phi_1(k_1)$ with $k_1>k^{*}_1$, the closed-loop system \eqref{eq-closed_loop_system_compacted_form} is asymptotically stable
for all $w\in\WW$.
Therefore, the robust stabilization problem of the augmented system \eqref{eq-augmented_system_transformed} is solved by the static state feedback control law \eqref{eq-control_law_augmented_system}.
\end{Proof}

\section{Proof of Lemma \ref{Result_of_output_feedback} }

\begin{Proof}
	Let $\bar\xi=\mathrm{col}\left(\bar\xi_1,\dots,\bar\xi_r\right)$.
	The closed-loop system composed of the augmented system \eqref{eq-transformed-augmented-system} and the control law \eqref{eq-output_control_law_augmented_system} is 
	\begin{align}\label{eq-closed_loop_augmented_system_OF}
		\dot{\bar z} & = A_1(w) \bar z+A_2(w) \bar{\xi}_1 \notag\\
		\dot{\bar\xi}&=A_4(w)\bar z+A_5(w)\bar\xi +B(w)\bar x_1 + B(w)\Psi T^{-1}_1\bar\eta_1  \notag\\
		\dot{\bar x}_1 & = \left(a-\Psi T^{-1}_1N_1\right) \bar x_1+\Psi T^{-1}_1 \left(aI_l-M_1-N_1\Psi T^{-1}_1\right)\bar\eta_1 \notag\\
		&\quad+b_a\Psi T^{-1}_2\bar\eta_2- b_a k_2 \left(\bar x_1+k_1\Gamma_1\varsigma\right) \notag\\
		\dot{\bar\eta}_1 & = \left(M_1+N_1\Psi T^{-1}_1\right) \bar\eta_1+N_1 \bar x_1  \notag\\
		\dot{\bar\eta}_2 & = \left(M_2+N_2\Psi T^{-1}_2\right) \bar\eta_2-k_2N_2(\bar x_1+k_1\Gamma_1\varsigma) \notag\\
		\dot\varsigma&=A_0(h)\varsigma+B_0(h)e 
	\end{align}
where
	\begin{align*}
		A_4(w)&=\begin{bmatrix}
			\zero \\
			A_3(w)
		\end{bmatrix}, \qquad  
B(w)=\begin{bmatrix}
			\zero \\
            b(w)
		\end{bmatrix} \\
		 A_5(w)&=\begin{bmatrix}
			0 & 1 & 0 & \cdots & 0 \\
			0 & 0 & 1 & \cdots & 0 \\
			\vdots & \vdots & \vdots & \ddots & \vdots \\
			0 & 0 & 0 & \cdots & 1 \\
			c_1(w) & c_2(w) & c_3(w) & \cdots & c_r(w)
		\end{bmatrix}   \\
	\Gamma_1&=\begin{bmatrix}
	            \gamma_0 & \gamma_1 & \cdots & \gamma_{r-2} & 1 
	          \end{bmatrix}.
	\end{align*}

From \eqref{eq-transformed-augmented-system}, we know that
	\begin{align*}
		e=\bar\xi_1, \quad	e^{(s)}=\bar\xi_{s+1},\quad s=1,\dots, r-1.
	\end{align*}
Hence, from \eqref{eq-control_law_augmented_system}, we have
	\begin{align}\label{eq-zeta1}
		\zeta_1=\Gamma_1\bar\xi.
	\end{align}
Similarly, from \eqref{eq-output_control_law_augmented_system}, we have
	\begin{align}\label{eq-hat-zeta1}
		\hat \zeta_1=\Gamma_1 \varsigma.
	\end{align}
Let $$\psi_s=h^{r-s}\left(e^{(s-1)}-\varsigma_s\right), \quad s=1, \ldots, r$$ 
and form $\psi=\mathrm{col}\left(\psi_1,\dots,\psi_r\right)$. 
From \eqref{eq-zeta1} and \eqref{eq-hat-zeta1}, 
we have
	\begin{align}\label{eq-error-zeta1-hatzeta1}
		\zeta_1-\hat\zeta_1=\Gamma_1 (\bar\xi - \varsigma)=\Gamma_1 D^{-1}_h\psi
	\end{align}
	where 
	\begin{align}\label{eq-Dh-def}
		D_h=\mathrm{diag}\left\{h^{r-1}, h^{r-2}, \ldots, 1\right\}. 
	\end{align}
	Then, it follows from \eqref{eq-closed_loop_augmented_system_OF} to \eqref{eq-error-zeta1-hatzeta1} that
\begin{align}
		\dot{\bar x}_1 
		& = \left(a-\Psi T^{-1}_1N_1\right) \bar x_1+\Psi T^{-1}_1 \left(aI_l-M_1-N_1\Psi T^{-1}_1\right)\bar\eta_1\notag\\
		&\quad+b_a\Psi T^{-1}_2\bar\eta_2-b_ak_2(\bar x_1+k_1\zeta_1)+b_ak_1k_2(\zeta_1-\hat\zeta_1) \notag\\
		& = \left(a-b_ak_2-\Psi T^{-1}_1N_1\right) \bar x_1\notag\\
& \quad +\Psi T^{-1}_1 \left(aI_l-M_1-N_1\Psi T^{-1}_1\right)\bar\eta_1+b_a\Psi T^{-1}_2\bar\eta_2 \notag\\
		&\quad-b_ak_1k_2\Gamma_1\bar\xi+b_ak_1k_2\Gamma_1 D^{-1}_h\psi \label{eq-dot-bar-x-1}\\
		\dot{\bar\eta}_2 & = \left(M_2+N_2\Psi T^{-1}_2\right) \bar\eta_2\notag\\
		&\quad -k_2N_2(\bar x_1+k_1\zeta_1)+k_1k_2N_2(\zeta_1-\hat\zeta_1)  \notag \\
		& = \left(M_2+N_2\Psi T^{-1}_2\right) \bar\eta_2-k_2N_2\bar x_1\notag\\
		&\quad -k_1k_2N_2\Gamma_1\bar\xi+k_1k_2N_2\Gamma_1 D^{-1}_h\psi. \label{eq-dot-bar-eta-2}
	\end{align}
	Moreover, with some routine computations, we can obtain
	\begin{align}\label{eq-dot-psi}
		\dot{\psi}_s & =h\left(-\delta_{r-s} \psi_1+ \psi_{s+1}\right), \quad s=1,\dots,r-1 \notag\\
		\dot{\psi}_r & =-h \delta_0 \psi_1+\dot{\bar\xi}_r 
	\end{align}
	where
	\begin{align*}
		\dot{\bar\xi}_r = A_3(w) \bar{z}+\sum_{s=1}^r c_s(w) \bar{\xi}_s+b(w)\bar x_1+b(w)\Psi T^{-1}_1\bar\eta_1.
	\end{align*}

	Let $\tilde X=\mathrm{col}\left(\bar z, \bar\xi, \bar x_1, \bar\eta_1, \bar\eta_2\right)$. 
Then, using \eqref{eq-dot-bar-x-1} to \eqref{eq-dot-psi}, 
the closed-loop system \eqref{eq-closed_loop_augmented_system_OF} can be compactly rewritten as
	\begin{align}\label{eq=closed-loop-system-OF}
		\begin{bmatrix} \dot{\tilde X}  \\ \dot\psi	\end{bmatrix}=
		\begin{bmatrix}
			\tilde A_{11}(w,k_1,k_2) & \tilde A_{12}(w,k_1,k_2,h)  \\
			\tilde A_{21}(w,k_1,k_2) &  h A_0(1)
		\end{bmatrix}
		\begin{bmatrix} \tilde X  \\  \psi \end{bmatrix}
	\end{align}
where 
	\begin{align}
		&\tilde A_{12}(w,k_1,k_2,h)
		=\begin{bmatrix}
			\zero \\ \zero \\ b_ak_1k_2  \\ \zero \\ N_2k_1k_2
		\end{bmatrix}\Gamma_1  D^{-1}_h  \label{eq-tildeA12-def}\\
	&\tilde A_{21}(w,k_1,k_2)
		=	\left[\begin{array}{ccccc}
			\zero & \zero & \zero & \zero & \zero \\
			A_3(w) & c(w) & b(w) & b(w)\Psi T^{-1}_1 & \zero
		\end{array}\right] \label{eq-tildeA21-def}	
	\end{align}
and $\tilde A_{11}(w,k_1,k_2)$ is defined in \eqref{eq-tildeA11-def},
	\begin{figure*}
		\hrulefill
			\begin{align}\label{eq-tildeA11-def}
				\tilde A_{11}(w,k_1,k_2)&=
				\begin{bmatrix}
					A_1(w) & A_2(w)D & \zero & \zero & \zero\\
					A_4(w) & A_5(w) & B(w) & B(w)\Psi T^{-1}_1 & \zero \\
					\zero & -b_ak_1k_2\Gamma_1 & C_1 & C_2 & b_a\Psi T^{-1}_2  \\
					\zero & \zero & N_1 & M_1+N_1\Psi T^{-1}_1  & \zero \\
					\zero & -k_1k_2N_2\Gamma_1 & -k_2N_2 & \zero & M_2+N_2\Psi T^{-1}_2
				\end{bmatrix} 
			\end{align}
		\hrulefill
	\end{figure*}
in which,
$C_1=a-b_ak_2-\Psi T^{-1}_1N_1$, 
$C_2=\Psi T^{-1}_1\left(aI_l-M_1-N_1\Psi T^{-1}_1\right)$, 
and $c(w)=\begin{bmatrix}
            c_1(w) & \cdots & c_r(w) 
          \end{bmatrix}$.
	
It is noted that the matrix $\tilde A_{11}(w,k_1,k_2)$ in \eqref{eq-tildeA11-def}
is similar to the matrix $\bar A(w,k_1,k_2)$ in \eqref{eq-bar-A-w-k-1-2-def}
through elementary transformations, which can be derived from \eqref{eq-transformation-2}. Since, by Lemma~\ref{Result_of_state_feedback}, there exist a positive constant $k_1^*$ and a positive function $\phi_1(k_1)$ such that for any $k_2>\phi_1(k_1)$ with $k_1>k^{*}_1$, the matrix $\bar A(w,k_1,k_2)$ is Hurwitz for all $w\in\WW$, the matrix $\tilde A_{11}(w,k_1,k_2)$ is also Hurwitz for all $w\in\WW$ under the same conditions.

From the definition of $D_h$ in \eqref{eq-Dh-def}, 
we know that $\Vert D^{-1}_h \Vert\leq 1$ for any $h\geq 1$. 
Thus, when $k_2>\phi_1(k_1)$ with a fixed $k_1>k^{*}_1$ 
and $h\ge 1$, there exist positive functions $\varphi_{i}(k_2), i=1,2$, 
satisfying $\Vert \tilde A_{12}(w,k_1,k_2,h) \Vert \leq \varphi_1(k_2)$ 
and $\Vert \tilde A_{21}(w,k_1,k_2)\Vert\leq \varphi_2(k_2)$ for all $w\in\WW$.
Then, by Lemma~\ref{Lemma_asymptotically_stable_system}, 
there further exists a positive function $\phi_2(k_2)$ such that, 
for any $h>\max\{1, \phi_2(k_2)\}$ and any $k_2>\phi_1(k_1)$ with $k_1>k^{*}_1$, 
system \eqref{eq=closed-loop-system-OF} is asymptotically stable for all $w\in\WW$.
Thus, the robust stabilization problem of the augmented system \eqref{eq-transformed-augmented-system} is solved by the  dynamic output feedback control law \eqref{eq-output_control_law_augmented_system}.
\end{Proof}

\end{document}